\documentclass{aa}
\usepackage{txfonts}
\usepackage{graphicx,natbib,bm,url,color,colortbl, xcolor}
\usepackage{amsmath}
\usepackage{txfonts}
\newcommand{\EQ}{\begin{equation}}
\newcommand{\EN}{\end{equation}}
\newcommand{\Eq}[1]{Eq.~(\ref{#1})}
\newcommand{\Fig}[1]{Fig.~\ref{#1}}
\newcommand{\Tab}[1]{Table~\ref{#1}}
\def\urms{u_{\rm rms}}
\graphicspath{{./fig/}{./png/}}

\title{Coagulation of inertial particles in supersonic turbulence}
\author{
Xiang-Yu Li \inst{1,2}
\and
Lars Mattsson \inst{1}
}

\institute{Nordita, KTH Royal Institute of Technology and Stockholm University,
10691 Stockholm, Sweden\\
\email{xiang.yu.li.phy@gmail.com}
\and
Section for Meteorology and Oceanography, Department of Geosciences, University of Oslo, P.O. Box 1022 Blindern,
0315, Oslo, Norway 
}
\date{\today}

\begin{document}

\abstract{Coagulation driven by supersonic turbulence is primarily an astrophysical problem because coagulation processes on Earth are normally associated with incompressible fluid flows at low Mach numbers, while dust aggregation in the interstellar medium (ISM) for instance is an example of the opposite regime. We study coagulation of inertial particles in compressible turbulence using high-resolution direct and shock-capturing numerical simulations with a wide range of Mach numbers from nearly incompressible to moderately supersonic. 
The particle dynamics is simulated by representative particles and the effects on the size distribution and coagulation rate due to increasing Mach number is explored. We show that the time evolution of particle size distribution mainly depends on the compressibility (Mach number).
We find that the average coagulation kernel $\langle C_{ij}\rangle$ scales linearly with the average Mach number
$\mathcal{M}_{\rm rms}$ multiplied by the combined size of the colliding particles,
that is, $\langle C_{ij}\rangle \sim \langle (a_i + a_j)^3\rangle\, \mathcal{M}_{\rm rms}\tau_\eta^{-1}$,
which is qualitatively consistent with expectations from analytical estimates.
A quantitative correction $\langle C_{ij}\rangle \sim \langle(a_i + a_j)^3\rangle(v_{\rm p,rms}/c_{\rm s})\tau_\eta^{-1}$
is proposed and can serve as a benchmark for future studies.
We argue that the coagulation rate $\langle R_c\rangle$ is also enhanced
by compressibility-induced compaction of particles.
}


\maketitle

\section{Introduction}
\label{sec:intro}
The kinetics of inertial particles that are (finite-size particles that are massive
enough to have significant inertia) in
turbulence has drawn much attentions for decades. It has been driven by wide applications in astrophysics, atmospheric sciences, and engineering.
The preferential concentration and fractal clustering of inertial particles in (nearly) incompressible turbulence has been simulated extensively \citep[see][]{Maxey87,Squires91,Eaton94,Bec03,Bec05,Bec07,Bec07b,Bhatnagar18,Yavuz18}.
In combination with the theory of coagulation of particles,
this has an important application in planet-formation theory 
\citep[see e.g.][]{Pan11,birnstiel2016dust, Johansen_2012,johansen2017forming}.
However, proto-planetary discs are dominated by low Mach-number
turbulence, which is not the case in many other astrophysical environments. One example is the cold phase of the interstellar medium (ISM), where turbulence is highly compressible with Mach numbers of order 10 and thus is dominated by shock waves. Only a few studies of inertial particles in high Mach-number turbulence can be found in the literature \citep[e.g.][]{Hopkins16,Mattsson19a,Lars19_Small},
and direct numerical simulations of turbulence-driven coagulation of inertial particles have not been performed so far.
Exploring the effects of compressibility (high Mach numbers) on coagulation is therefore an important branch of research that is now becoming possible through the rapid development of computing power.

From an astrophysical perspective, cosmic dust grains, which are made of mineral or carbonaceous material and are a perfect example of inertial particles, are ubiquitous throughout the universe. Rapid interstellar dust growth by accretion of molecules is thought to be necessary to compensate for various dust destruction processes \citep[see e.g.][]{Mattsson11b,Valiante11,Rowlands14}. Grains may also grow by aggregation or coagulation (which does not increase the dust mass), however, when the growth by accretion has produced large enough grains for coagulation to become efficient, that is, once the `coagulation bottleneck' has been passed \citep{Mattsson16}.
How efficient the coagulation is in molecular clouds (MCs) is not fully understood, although models and simulations have suggested that turbulence is the key to high growth efficiency \citep{elmegreen2004interstellar,Hirashita09,Hirashita10, pan2014turbulenceII,pan2014turbulenceIII,pan2015turbulence,Hopkins16,Mattsson19a,Lars19_Small} and observations indicate the presence of very large grains (which can be tens of $\mu$m across) in the cores of MCs \citep{Hirashita14}. 

We aim to study the role that the Mach number plays for the coagulation of inertial particles in compressible turbulence. Specifically, we strive to study how coagulation of particles depends on Mach number in regions where the particles cluster and form filaments. The purpose is not to target any specific astrophysical context, but to explore the Mach-number dependence to the extent that this is computationally feasible. 
There are three main challenges. First, the dynamics of inertial particles in {\it \textup{compressible}} turbulence is poorly understood. Second, the coagulation process is a non-equilibrium process, as the particle size distribution evolves with time. Third, the coagulation timescale and the characteristic timescale of the turbulence are very different in dilute systems, such as those typically studied in astrophysics. 
In classical kinetic theory, the collision kernel $C_{ij}$ is a function of the relative velocity $\Delta \bm{v}_{ij}$ of two particles $i$ and $j$, which is difficult to calculate analytically, except in some special cases. For the same reason, it is difficult to calculate the coagulation rate using analytical models. More exactly, the classical \citet{Smoluchowski16} problem has only three known exact solutions \citep{Aldous99}, and numerical solution of the coagulation equation is only feasible if treated as a local or `zero-dimensional' problem.

The main objective of the present work is to offer a way to quantify and possibly parametrise the effects of turbulent gas dynamics and hydrodynamic drag on the coagulation rate in such a way that it can be included for instance in traditional models of galactic chemical evolution (including dust), which are based on average physical quantities \citep{Mattsson16}. 
A major problem when simulating the dust growth in the ISM is that
the system is large scale and dilute. The coagulation rate is extremely low in such a system, which leads to very different timescales for the turbulent gas dynamics and coagulation. 

\section{Turbulence and kinetic drag}

In this section, equations governing compressible flow and
particle dynamics of
inertial particles (e.g.\ dust grains) are presented. The 
{\sc Pencil Code} with HDF5 IO \citep{brandenburg2020pencil, 2020GApFD}
is used to solve these equations.

Since the carrier fluid in our study is isothermal,  its turbulence described by \Eq{turb} is scale free, that is, the box size $L$, the mean mass density $\langle\rho\rangle$, and the sound speed $c_{\rm s}$ are the unit length, unit density, and unit velocity, respectively. These quantities can thus be scaled freely. However, the inclusion of coagulation process means our simulation is no longer scale free. This requires a careful treatment of initial conditions and scaling of units, which is discussed in more detail in Section~\ref{sec:initial}.

\subsection{Momentum equation of the carrier flow}
\label{sec:flow}
The motion of the gas flow is governed by
the Navier-Stokes equation,
\begin{equation}
        {\partial{\bm u}\over\partial t}+\bm{u}\cdot{\bm{\nabla}}\bm{u}={\bm f}
-\rho^{-1}{\bm{\nabla}} p
        +\rho^{-1}\bm{F}_{\rm visc} ,
\label{turb}
\end{equation}
where ${\bm f}$ is a forcing function \citep{Brandenburg01},
$p$ is the gas pressure, and $\rho$ is the fluid or gas density obeying
the continuity equation,
\begin{equation}
        {\partial\rho\over\partial t}+{\bm{\nabla}}\cdot(\rho\bm{u})=0 .
\end{equation}
For the case of direct numerical simulation with a constant kinetic viscosity of the gas
flow, the viscosity term $\bm{F}_{\rm visc}$ equals the physical
viscosity term $\bm{F}_{\rm visc}^{\nu}$ given by
\begin{equation}
\bm{F}_{\rm visc}^{\nu}=
 \rho \nu\left({\bm\nabla}^2\bm{u}+\frac{1}{3}\nabla\nabla\cdot\bm{u}+2\bm{{\sf S}}\cdot\nabla\ln\rho\right),
\end{equation}
where $\bm{{\sf S}}={1\over 2}\left[{\bm\nabla} \bm{u} +\left({\bm\nabla} \bm{u} \right)^{T}\right]-{1\over 3}\left({\bm\nabla} \cdot \bm{u} \right){\sf I}$
is the rate-of-strain tensor (${\sf I}$ is the unit tensor)
resulting from the shear effect and the deformation
due to compression.
For the case with shock-capturing viscosity, the viscosity term becomes
\begin{equation}
  \bm{F}_{\rm visc}=
  \bm{F}_{\rm visc}^{\nu}
  +\rho\zeta_{\rm shock}\nabla\nabla\cdot\bm{u}+\left(\nabla\cdot\bm{u}\right)\nabla\left(\rho\zeta_{\rm shock}\right).
\end{equation}
The shock viscosity $\zeta_{\rm shock}$ is given by 
\begin{equation}
  \zeta_{\rm shock}=c_{\rm shock}\langle\rm{max}[(-{\bm\nabla}\cdot\bm{u})_+]\rangle(\min(\delta x,\delta y,\delta z))^2,
\end{equation}
where $c_{\rm shock}$ is a constant defining the strength of the shock viscosity \citep{haugen2004mach}.
The length of the lattice is given by $\delta x$, $\delta y$, and $\delta z$, respectively.
$\bm{F}_{\rm visc}^{\rm shock}$ is used in simulations with high Mach number,
where we strive to use the highest spatial resolution to capture the shocks.
Nevertheless, it is necessary to introduce this term to handle the strongest
shocks.
Two dimensionless parameters characterise compressible turbulence: the Reynolds number Re and the root-mean-square (rms) Mach number $\mathcal{M}_{\rm rms}$. Re is defined as
\EQ
{\rm Re} \equiv u_{\rm rms}L_{\rm inj}/\nu,
\label{eq:Re}
\EN
where $u_{\rm rms}$ is the rms turbulent velocity
and $L_{\rm inj}$ is the energy injection length scale.
The compressibility of the flow is characterised by $\mathcal{M}_{\rm rms}$, which is defined as
\EQ
{\cal{M}_{\rm rms}}=\urms/c_{\rm s},
\label{eq:Ma}
\EN
where $c_{\rm s}$ is the sound speed. The sound speed
is kept constant because the compressible flow to be investigated
here is assumed to be isothermal such that $c_{\rm s}^2=\gamma p/\rho$,
where $\gamma=c_{\rm P}/c_{\rm v}=1$ with the specific heats $c_{\rm P}$ and $c_{\rm V}$
at constant pressure and constant volume, respectively.
Another quantity is the mean energy dissipation rate $\langle\bar{\epsilon}\rangle$, which measures how vigorous the small eddies are in turbulence. It can be calculated from the trace of $\bm{{\sf S}}_{ij}$ as $\langle\bar{\epsilon}\rangle=2\nu\, \textstyle{\overline{{\rm Tr\,} {\sf S_{ij}} {\sf S_{ji}}}}$. $\langle\bar{\epsilon}\rangle$ determines the smallest scales of the turbulence, for example, the Kolmogorov length scale is defined as $\eta=(\nu^3/{\langle\bar{\epsilon}\rangle})^{1/4}$ and the timescale is defined as $\tau_{\eta}=(\nu/\langle\bar{\epsilon}\rangle)^{1/2}$.
Becausee the Saffman-Turner collision rate is proportional
to $\tau_\eta^{-1}$ \citep{1955_Saffman}, $\langle\bar{\epsilon}\rangle$ indirectly determines the coagulation rate of particles in an incompressible flow.
Coagulation occurs at the small scales of turbulence, and the strength of the small eddies determines the particle velocity \citep{li2017effect}. Therefore it is worth investigating  whether and how
$\langle\bar{\epsilon}\rangle$ affects the coagulation rate
in compressible turbulence as well. We show
that it {\em \textup{does not}} affect the coagulation rate in
the compressible case. 

The stochastic solenoidal forcing $\bm{f}$ is given by
\begin{equation}
  \bm{f}(\bm{x},t)=\mbox{Re}\{N\bm{f}_{\bm{k}(t)}\exp[i\bm{k}(t)\cdot\bm{x}+i\phi(t)]\},
\end{equation}
where $\bm{k}(t)$ is the wave space,
$\bm{x}$ is position, and $\phi(t)$ ($|\phi|<\pi$) is
a random phase. The normalization factor
is given by $N=f_0 c_{\rm s}(kc_{\rm s}/\Delta t)^{1/2}$, where $f_0$ is a
non-dimensional factor, $k=|\bm{k}|$, and $\Delta t$ is the
integration time step \citep{BD02}.
We chose a completely non-helical
forcing, that is,
\begin{equation}
  \bm{f}_{\bm{k}}=\left(\bm{k}\times\bm{e}\right)/\sqrt{\bm{k}^2-(\bm{k}\cdot\bm{e})^2},
\end{equation}
where $\bm{e}$ is the unit vector.

To achieve different $\cal{M}_{\rm rms}$ with fixed ${\rm Re}$ and $\langle\bar{\epsilon}\rangle$
in the simulations, we need to change $u_{\rm rms}$, $\nu$, and the simulation box $L$ simultaneously
according to \Eq{eq:Re} and \Eq{eq:Ma},
and we also considered
\begin{equation}
  \langle\bar{\epsilon}\rangle \sim \frac{u_{\rm rms}^3}{L_{\rm inj}}.
  \label{eq:epsilon}
\end{equation}
Since $u_{\rm rms}$ is essentially determined by the amplitude of forcing $f_0$, we changed $f_0$ in the simulation as well.

\subsection{Particle dynamics}
The trajectories of inertial particles
is determined by
\begin{equation}
        \frac{d\bm{x}_i}{dt}=\bm{v}_i
\label{dxidt}
\end{equation}
and
\begin{equation}
        \frac{d\bm{v}_i}{dt}=\frac{1}{\tau_i}(\bm{u}
        -\bm{v}_i)\,,
\label{dVidt}
\end{equation}
where
\begin{equation}
\label{stoppingtime}
\tau_i = \sqrt{\pi\over 8}{\rho_{\rm mat}\over\rho}{a\over  c_{\rm s}} \left(1 + {9\pi\over 128}{|\bm{u}-\bm{v}_i|^2\over c_{\rm s}^2 } \right)^{-1/2},
\end{equation}
is the stopping time, that is, the kinetic-drag timescale.
In the equation above, $a$ is the radius of a particle, $\rho_{\rm mat}$
is the material density of particles, and $\rho$ is the mass density of the gas.
We assumed that particles are well described in the Epstein limit
because the mean-free-path $\lambda$ is large and particles are small
in most astrophysical contexts \citep[large Knudsen number, Kn~$= \lambda/a \gg 1$][]{armitage2010astrophysics}. 
The stopping time at low relative Mach number  ($\mathcal{W} = |\bm{u}-\bm{v}_i|/c_{\rm s}\ll1$) is
\begin{equation}
\tau_i (\mathcal{W}\ll 1) = \sqrt{\pi\over 8}{\rho_{\rm mat}\over\rho}{a\over c_{\rm s}}
.\end{equation}
The term in parentheses of \Eq{stoppingtime} is a correction for high $\mathcal{W}$. \Eq{stoppingtime} is essentially a quadratic interpolation between the two expressions for the limits $\mathcal{W}\ll 1$ and $\mathcal{W}\gg 1$ derived from the exact expression for $\tau_i$ \citep[see][]{Schaaf63,Kwok75,Draine79}.

To characterize the inertia of particles, we define a `grain-size parameter' as
\begin{equation}
\label{eq:alphadef}
\alpha = {\rho_{\rm mat}\over\langle \rho\rangle}{a\over L} ,
\end{equation}
which is the parametrisation used  by \citet{Hopkins16}.
Because the total mass of a simulation box of size $L$,
as well as the mass of a grain of a given radius $a$, is constant,
the quantity $\alpha$ is solely determined by $a$ regardless of
the characteristics of the simulated flow.

In general, the inertia of particles is characterised by the Stokes number ${\rm St}=\tau_i/\tau_\eta$.
The disadvantage of ${\rm St}$ as the size parameter for inertial particles in a highly compressible carrier fluid is that a fluid flow with ${\rm Re} \gg 1$ cannot be regarded as a Stokes flow. If $\mathcal{M}_{\rm rms}\gg 1$ as well, ${\rm St}$ is not even well defined as an average quantity in a finite simulation domain. The parameter $\alpha$ is therefore a better dimensionless measure of grain size than the average Stokes number $\langle {\rm St}\rangle$ for a
supersonic compressible flow. Moreover,  $\langle {\rm St}\rangle$ is not only a function of the size, but also a function of the mean energy dissipation rate $\langle\bar{\epsilon}\rangle$, which complicates the picture even further.

\subsection{Averages}
In the following we frequently refer to the mean or average quantities of three different types. For each of them we use a different notation. First, we use the bracket notation $\langle Q\rangle$ for volume averages, taken over the whole simulation box unless stated otherwise. Second, we use the over-bar notion $\bar{Q}$ for straight time-averaged quantities. Third, we use the tilde notation $\tilde{Q}$ for ensemble averages, that is, averages defined by the first moment of the distribution function of the particles.

The rms value of a fluctuating physical quantity has been mentioned in section \ref{sec:intro}. In terms of the above notion, rms values always refer to $Q_{\rm rms} \equiv \sqrt{\langle Q^2 \rangle}$.

\section{Coagulation}

%XY: removed as suggested.
%Coagulation\LEt{a single sentence does not constitute a paragraph. Please either add to this or remove} algorithm of inertial particles and theoretical models
%are presented in this section.
%XY.

\subsection{Numerical treatment of coagulation}
\label{method:sp}
The most physically consistent way to model coagulation is to track each individual
Lagrangian particle and to measure the collisions among them when they
overlap in space, which is computationally challenging because the coagulation timescale of inertial particles is often much shorter than the Kolmogorov timescale. We also used $10^7$ representative particles, which means solving a large $N$-body problem. Because of the aforementioned computational load, a super-particle approach is often used to study the coagulation of dust grains \citep{Dullemond_2008,Johansen_2012,li2017eulerian}. Instead of tracking each individual particles, super-particles consisting of several identical particles are followed. Within each super-particle, all the particles have the same velocity $\bm{v}_i$ and size $a$. The super-particle approach is a Monte Carlo approach, which treats coagulation of dust grains in a stochastic manner \citep{bird1978monte,bird1981monte,jorgensen1983comparison}. Each super-particle is assigned a virtual volume that is the same as the volume of the lattice, therefore a number density
$n_j$.

When two super-particles $i$ and $j$ reside in the same grid cell, the probability of coagulation is $p_{\rm c}=\tau_{\rm c}^{-1}\Delta t$, where $\tau_{\rm c}$ is the coagulation time and $\Delta t$ is the integration time step. A coagulation event occurs when $p_{\rm c}>\eta_c$, where $\eta_c$ is a random number. The coagulation timescale $\tau_{\rm c}$ is defined as
\begin{equation}
        \tau_{\rm c}^{-1}=\sigma_{\rm c} n_j\,|{\bm w}_{ij}|\, E_{\rm c},
\label{tauij1}
\end{equation}
where $\sigma_{\rm c}=\pi(a_i+a_j)^2$  and ${\bm w}_{ij}$ are the geometric coagulation cross section and the absolute velocity difference between two particles with radii $a_i$  and $a_j$, respectively, and $E_{\rm c}$ is the coagulation efficiency \citep{klett1973theoretical}.
For simplicity, we set $E_{\rm c}$ to unity.
This means that all particles coalesce upon collision, that is, bouncing and fragmentation are neglected.
This treatment may overestimate the collision rate but does not affect the dynamics of particles.
Therefore the $\mathcal{M}$ dependence should not be affected.
Compared with the super-particle algorithm that is widely used in planet formation \citep{Dullemond_2008, Johansen_2012},
our algorithm provides better collision statistics \citep{li2017eulerian}.
We refer to \citet{li2017eulerian}
for a detailed comparison of the super-particle algorithm used in \citet{Johansen_2012} and \citet{li2017eulerian,li2017effect,Li19}.

\subsection{Timescale differences}
\label{sec:timescales}
Before we describe the basic theory of coagulation of particles in a turbulent carrier fluid,
it is important that we consider the different timescales that are involved in this complex and composite problem.
In \Eq{tauij1}, we introduced $\tau_{\rm c}$.
The other important timescale in a model of coagulation of particles in turbulence is the flow timescale of the carrier fluid, in this case, the large-eddy turnover time
$\tau_L= L/u_{\rm rms}$,
where $u_{\rm rms}$ is the rms flow velocity.
Clearly, $\tau_L$ depends on the scaling of the simulation. When we compare the two timescales, we
find that
\begin{equation}
  {\tau_L \over \langle{\tau_{\rm c}\rangle}} \sim N_{\rm p}{a^2\over L^2}{\langle{|\bm{w}_{ij}|}\rangle\over u_{\rm rms}},   
\end{equation}
where $N_{\rm p}$ is the total number of particles in the simulated volume.
We note that $\langle|{\bm w}_{ij}|\rangle / u_{\rm rms} \ll 1$ as long as
the particles do not decouple completely from the carrier flow.
In order to avoid slowing down the simulation too much,
we aim for $\tau_L / \langle\tau_{\rm c}\rangle\sim 1$.
From this we may conclude that $N_{\rm p} \gg (L/a)^2$, which implies that if we have an upper bound of $N_{\rm p}$ for computational reasons, we cannot simulate tiny particles in a large volume.
The ratio $\tau_L/\langle{\tau_{\rm c}\rangle}$ shows how difficult it can be to simulate the coagulation in astrophysical contexts, in particular when the details of coagulation of inertial particles are simulated in a carrier fluid representing well-resolved compressible turbulence. 

In addition to the two timescales discussed above, we must also consider the stopping time $\tau_i$ of the particles because we study inertial particles. For $\alpha \lesssim 0.1$, $\tau_i$ is typically smaller than $\tau_L$. Hence, the competing timescales would rather be $\tau_{\rm c}$ and $\tau_i$, which suggests that the ratio $\tau_i/\tau_{\rm c}$ should be of order unity to avoid slowing down the simulation compared to the case of non-interacting particles. By the same assumptions as above ($k_{\rm f}\approx 3$ and $E_{\rm c} \sim 1$), we can show that 
\begin{equation}
  {\tau_i \over \langle{\tau_{\rm c}\rangle}} \sim 
  {\langle|\bm{w}_{ij}|\rangle\over c_{\rm s}}{\rho_{\rm p}\over \rho}\,a_i^3
,\end{equation}
where $\rho_{\rm p}{\equiv\rho_{\rm mat}\,n_i}$ is the mass density of particles (not to be confused with the bulk material density $\rho_{\rm mat}$). In many astrophysical contexts (in particular, cold environments) $\langle|\bm{w}_{ij}|\rangle /c_{\rm s}\sim 1$, which then suggests we must have $\rho_{\rm p}/ \rho \sim 1$. This is always inconsistent with cosmic dust abundances, however, whether in stars, interstellar clouds, or even proto-planetary discs. In the cold ISM, $\rho_{\rm p}/ \rho \sim 10^{-2}$ and $\langle|\bm{w}_{ij}|\rangle /c_{\rm s}\sim 1$,
which implies that $\tau_i/\tau_{\rm c}\ll 1$
and thus the time step of a simulation of coagulation in such an environment
is limited by $\tau_i$.
In practice, this means that it will be difficult (or even impossible) to target coagulation in cold molecular clouds in the ISM without highly specialised numerical methods.

The goal of the present study is primarily to investigate how coagulation of inertial particles depends on $\mathcal{M}_{\rm rms}$ and not to simulate coagulation in a realistic and dilute astrophysical environment. We note, however, that any result  on coagulation of particles in compressible turbulence is primarily of importance for astrophysics, for instance, the processing of dust grains in the ISM and various types of circumstellar environments. Therefore we tried to make the simulation system as dilute, while still ensuring statistical convergence and computational feasibility.

\subsection{Theory of coagulation of inertial particles in turbulence}
\label{sec:theory}

Coagulation, as described by the \citet{Smoluchowski16} equation, is determined by the coagulation rate $R_{ij}$ between two grains species (sizes) $i$ and $j$ and the associated rates of depletion. In general, we have $R_{ij} = {1\over 2} n_i\,n_j\, C_{ij}$, where $n_i$, $n_j$ are the number densities of the grains $i$ and $j$, and $C_{ij}$ is the collision kernel.
Turbulence has been proposed to have a profound effect on $C_{ij}$,
and we focus this theory section on what happens to $C_{ij}$.

Assuming the distribution of particle pairs can be separated into distinct spatial and velocity  distributions, we have \citep{Sundaram97}
\begin{equation}
\label{kernel_gen}
    \langle C_{ij}\rangle = \pi (a_i+a_j)^2 \,g(r,a_i,a_j)\int\bm{w}_{ij}\,P(\bm{w}_{ij}, a_i, a_j)\,d\bm{w}_{ij},
\end{equation}
where $g$ is the radial distribution function (RDF) and $P$ is the probability density distribution of relative velocities $\bm{w}_{ij}$. 

Below, we review the basic theory of coagulation in the tracer particle limit and large-inertial limit.
Since small-clustering is negligible for both small- and large-inertial particles, we implicitly assumed that $g(r)=1$.
Moreover, in case of Maxwellian velocities,
that is, $P(\bm{w}_{ij}, a_i, a_j)$ follows a Maxwellian distribution,
the integral part of \Eq{kernel_gen} becomes $\langle \bm{w}_{ij} \rangle = \sqrt{8\pi/3\, \langle \bm{w}_{ij}\cdot\bm{w}_{ij} \rangle}$.
Thus, the collision kernel in \Eq{kernel_gen} reduces to
\begin{equation}
\langle C_{ij}\rangle = \sqrt{8\pi/3}\,(a_i+a_j)^2 \sqrt{\langle \bm{w}_{ij}\cdot \bm{w}_{ij}\rangle},
\end{equation}
which is the form assumed in the two following subsections.

\subsubsection{Tracer-particle limit}

In the low-inertial limit, also known as the Saffman-Turner limit or the tracer-particle limit \citep{1955_Saffman},
$\langle \bm{w}_{ij}^2 \rangle$ 
is a simple function of $a_i$ and $a_j$. In case of a mono-dispersed grain population ($a = a_i = a_j$) suspended in a turbulent low Mach-number medium, we may use the \citet{1955_Saffman} assumption, $\langle \bm{w}_{ij}^2 \rangle = {1\over 5} (a/\tau_\eta)^2$, where $\tau_\eta = \sqrt{\nu/\langle\bar{\epsilon}\rangle}$ is the Kolmogorov timescale. This is a reasonable approximation if $\mathcal{M}_{\rm rms}$ is not too large. The \citet{1955_Saffman} theory relies on $\bm{w}_{ij}$ having a Gaussian distribution (Maxwellian velocities) and the final expression for $\langle C_{ij}\rangle$ becomes
\begin{equation}
  \langle C_{i}\rangle =\sqrt{8\pi\over 15}\frac{\left(2a_i\right)^3}{\tau_\eta}. 
\label{eq:S-T}    
\end{equation}
For the multi-dispersed case, we replace $2a_i$ by $(a_i+a_j)$ in \Eq{eq:S-T}.

At first sight, compressibility does not seem to play any role at all, given the collision kernel $\langle C_{ij}\rangle$ above. However, it can affect the number density of particles $n_i$, and therefore, $R_{ij}$. In the tracer particle limit, the spatial distribution of particles is statistically the same as for gas. Simulations have shown that the gas density of isothermal hydrodynamic turbulence exhibits a lognormal distribution \citep[e.g.][]{Federrath09, Hopkins16, Mattsson19a} with a standard deviation of $\sigma_\rho^2$ that is empirically related to ${\cal M}_{\rm rms}$ \citep[see e.g.][]{Passot98,Price11,Federrath10}. Consequentially, $n_i$ depends on ${\cal M}_{\rm rms}$, and so does $R_{ij}$.

\subsubsection{Large-inertia limit}

In the opposite limit, the large-inertia limit, particles should behave according to kinetic theory.
As shown by \citet{Abrahamson75},
we have in this limit that particles are randomly positioned and follow a Maxwellian velocity distribution. In this case, we may conclude that $\langle \bm{w}_{ij}^2 \rangle = \langle \bm{v}_{i}^2 \rangle + \langle \bm{v}_{j}^2 \rangle$ because the particles are then statistically independent and thus they have a covariance that is identically zero. As in the tracer-particle limit,
the expression for a mono-disperesed population becomes
\begin{equation}
\label{eq:sameakin}
    \langle C_{i}\rangle = a^2 \left({16\pi\over 3} \langle \bm{v}_i^2\rangle \right)^{1/2}. 
\end{equation}
Previous theoretical work on inertial particles in turbulent flows \citep[e.g.][]{Abrahamson75,Hopkins16,Mattsson19a,Pan13,Wang00}
have shown that the rms velocity $v_{\rm rms}$ of the particles is a function of their size.
More precisely, we can parametrise in terms of grain size such that
$v_{\rm rms}^2(a_i)/u_{\rm rms}^2 \approx a_0/a_i$ for $a/a_0 \gg 1$, where $a_0$ is a scaling radius that can be estimated from the stopping time $\tau_i$ and the integral timescale $\tau_L$ \citep{Hedvall19,Mattsson21}. Assuming Maxwellian velocity distributions again, we have that $\langle \bm{w}_{ij}\cdot \bm{w}_{ij}\rangle  = v_{\rm rms}^2(a_i)+ v_{\rm rms}^2(a_j)$.
Thus, the mean collision kernel for the multi-dispersed case  can be expressed as
\begin{equation}
\langle C_{ij} \rangle \approx \sqrt{8\pi\over 3}\, (a_i+a_j)^2\,u_{\rm rms}\,\left({a_0\over a_i} +{a_0\over a_j}\right)^{1/2}.
\label{eq:cij}
\end{equation}
Here, we may note that $\langle C_{ij} \rangle$ is proportional to $u_{\rm rms}$, implying that the collision rate should scale with $\mathcal{M}_{\rm rms}$. For particles of equal size, that is, $a_i = a_j$, $\langle C_{ij} \rangle$ reduces to the form given above in Eq. (\ref{eq:sameakin}), which also means that $\langle C_{i} \rangle \sim a_i^{5/2}$. In the absence of external body forces other than kinetic drag, $\langle{\bm v}_i^2\rangle \to 0$ for very large inertia particles. Thus, $\langle C_{ij} \rangle \to 0$, eventually. 

To summarise, we note that for the tracer-particle limit (small-inertia particles, St~$\ll1$), density variations in high-${\cal M}_{\rm rms}$ turbulence, and the locally elevated number densities ($n_i$) of dust particles that follows consequently, can have a significant impact on the average collision rate $\langle R_{ij} \rangle$, although not on $\langle C_{ij} \rangle$.
For particles with large inertia, beyond the effect of compressibility on $\langle R_{ij} \rangle$, caustics in the particle phase may also contribute to $\langle C_{ij} \rangle$ (see Appendix \ref{app:kernel} for details of the discussion).

\subsection{Initial conditions}
\label{sec:initial}
Super-particles were initially distributed randomly in the simulation box and mono-dispersed in size ($\alpha_0=10^{-4}$). As discussed in section~\ref{method:sp}, each super-particle was assigned a virtual volume $(\delta x)^3$, where $\delta x$ is the lateral size of the lattice. With an initial number density of dust grains $n_0$, the total number of dust grains in the computational domain is given by
\EQ
n_0L^3=N_{\rm s}(n_j\delta x^3),
\label{eq:n0}
\EN
where $N_{\rm s}$ is initial total number of super-particles. Since $(L/\delta x)^3=N_{\rm grid}$ with $N_{\rm grid}$ the number of grid cells, \Eq{eq:n0} can be rewritten as
\EQ
n_0N_{\rm grid}=N_{\rm s}n_j,
\label{eq:ns}
\EN
where $n_j$ is the number density within each super-particle at $t = 0$. The number of physical particles in each super-particle $N_{\rm p/s}$ is determined by
\EQ
N_{\rm p/s}=N_{\rm p}/N_{\rm s}=\frac{L^3}{N_{\rm grid}}n_j,
\label{eq:Nps}
\EN
which means that $N_{\rm p/s}$ is uniquely determined by $L^3$ when $N_{\rm s}$, $N_{\rm grid}$, and $n_j$ are fixed.

To avoid running out of memory while executing the simulation, we must limit the number of super-particles to $N_{\rm s} \sim 10^7$, which leads to a required resolution $N_{\rm grid}=512^3$. The value of $n_0$ must be chosen for computational feasibility, and we also kept the number of particles within each super-particle to a minimum to avoid averaging out too much of the turbulence effects (as explained in Section \ref{method:sp}). 

We can take the physical parameters of dust grains in the ISM as an example of how difficult it is to simulate a dilute system, even on a current modern supercomputer. According to \Eq{tauij1} and \Eq{eq:n0}, the collision frequency is proportional to $n_0$. With $n_0=3.33\times10^{-7}\,\rm{cm}^{-3}$, $a_0=10^{-6}\,\rm{cm}$, $|\bm{v}_i-\bm{v}_j|\approx10^5\,\rm{cm\, s^{-1}}$, and $E_c\approx 1$, the initial collision frequency is $\tau_c^{-1}\approx 10^{-9}\,\rm{s}^{-1}$. The simulation time step must match the corresponding physical coagulation timescale for the particles, which is far beyond the computational power at our disposal in case of a small number of particles within each super-particle.

\subsection{Diagnostics}
\label{sec:dia}
The coagulation process is sensitive to the large-particle tail of the particle size distribution $f(a,t)$ because particle coagulation is strongly dependent on the total cross section. The tails of $f(a,t)$ can be characterised using the normalised moments of radius $a$
 \citep{li2017eulerian},
\begin{equation}
  a_\zeta=\left({M_\zeta}/{M_0}\right)^{1/\zeta},
\label{azeta}
\end{equation}
where 
\begin{equation}
M_\zeta(t)=\int_{0}^\infty f(a,t)\,a^\zeta \,{\rm d}a, 
\end{equation}
is the $\zeta$th moment of $a$. We adopted $\zeta=24$ to characterise the large-particle tail. 
To follow the overall evolution of $f(a,t)$, we also considered the mean radius, defined by the first-order normalised moment, $\tilde{a} = a_1 = M_1/M_0$.
The relative standard deviation of $f(a,t)$ can be defined as
\EQ
\sigma_a/\tilde{a}=(a_2^2-a_1^2)^{1/2}/a_1,
\EN
where $\sigma_a=(a_2^2-a_1^2)^{1/2}$ is the standard deviation of $a$. 

\section{Results}
\label{sec:results}

\begin{table*}
\caption{Parameter values used in different simulation runs.
The sound speed is $c_s=1.0\, [L/T]$ in all simulations except
for Run F, where $c_s=0.5\, [L/T]$.
}
\centering
\setlength{\tabcolsep}{3pt}
\begin{tabular}{lccccccccccccr}
  Run &  $f_0$ & $L_x$ & $N_{\rm grid}$ & $N_{\rm s}$& $\alpha_{\rm ini}$ & $\nu$& $c_{\rm shock}$& ${\cal M}_{\rm rms}$ & ${\rm Re}$& $\langle\bar{\epsilon}\rangle$& $\eta$  & $\rm{St}^L(t=0)$ & $\rm{St}^\eta(t=0)$\\
\hline
  A & $1.0$ & $2\pi$ & $512^3$ & $15624960$ & 0.016&$2.5\times10^{-3}$ & 2.0 &1.56 &200 & 0.80 & 0.012  & 0.047 & 1.071 \\ %3D/SW512condens0_coag_grav0_turb_nu2p5em3_helical
  C & $0.73$ & $7.47$ & $512^3$ & $15624960$& 0.013 & $2.5\times10^{-3}$ & 2.0 & 1.27 & 194  & 0.42 & 0.014  &0.032 & 0.769 \\ %3D/SW512condens0_coag_grav0_turb_nu2p5em3_helical_f0p7_L7
  D & $0.02$ & $0.05$ & $256^3$ & $1953120$& 2.000 & $1\times10^{-5}$& --   & 0.066 & 83 & 0.76 & 0.0002  & 0.200 & 2.778 \\ %3D/SW256condens0_coag_f0p088_L0p05
  E & $1.20$ & $0.16$ & $256^3$ & $1953120$& 0.625  & $5\times10^{-5}$&2.0 & 0.55  & 90  & 1.13 & 0.0007  & 0.667 & 10.000\\ %3D/SW256condens0_coag_nu5em5_f1p2_L0p16
  F & $3.50$ & $\pi$ & $512^3$ & $15624960$& 0.032 & $2.5\times10^{-3}$&8.0 & 2.58  & 82  & 1.03 & 0.01   & 0.167 & 2.600 \\ %3D/SW512condens0_coag_nu2p5em3_L3_f3_cs0p5_nuS8_256CPU
  G & $1.00$ & 1.60 & $256^3$ & $1953120$& 0.032    & $1\times10^{-3}$&2.0  & 0.99  & 81  & 0.80 & 0.006  & 0.118 & 1.500\\ %3D/SW256condens0_coag_grav0_turb_nu1em3_f1_l1p6
  H & $1.00$ & 0.60 & $256^3$ & $1953120$& 0.167    & $2.5\times10^{-4}$ &2.0& 0.75  & 92  & 0.79 & 0.002  & 0.240 & 3.000 \\ %3D/SW256condens0_coag_grav0_turb_nu2p5em4_f1_l0p6
  I & $1.00$ & 0.80 & $256^3$ & $1953120$& 0.125    & $1\times10^{-3}$ & 2.0 & 0.74  & 30  & 0.79 & 0.006  & 0.176 & 1.500\\ %3D/SW256condens0_coag_grav0_turb_nu1em3_f1_l0p8

\hline
\multicolumn{14}{p{0.5\textwidth}}{}
\end{tabular}
\label{tab:helical}
\end{table*}

\begin{figure}
\resizebox{\hsize}{!}{
\includegraphics{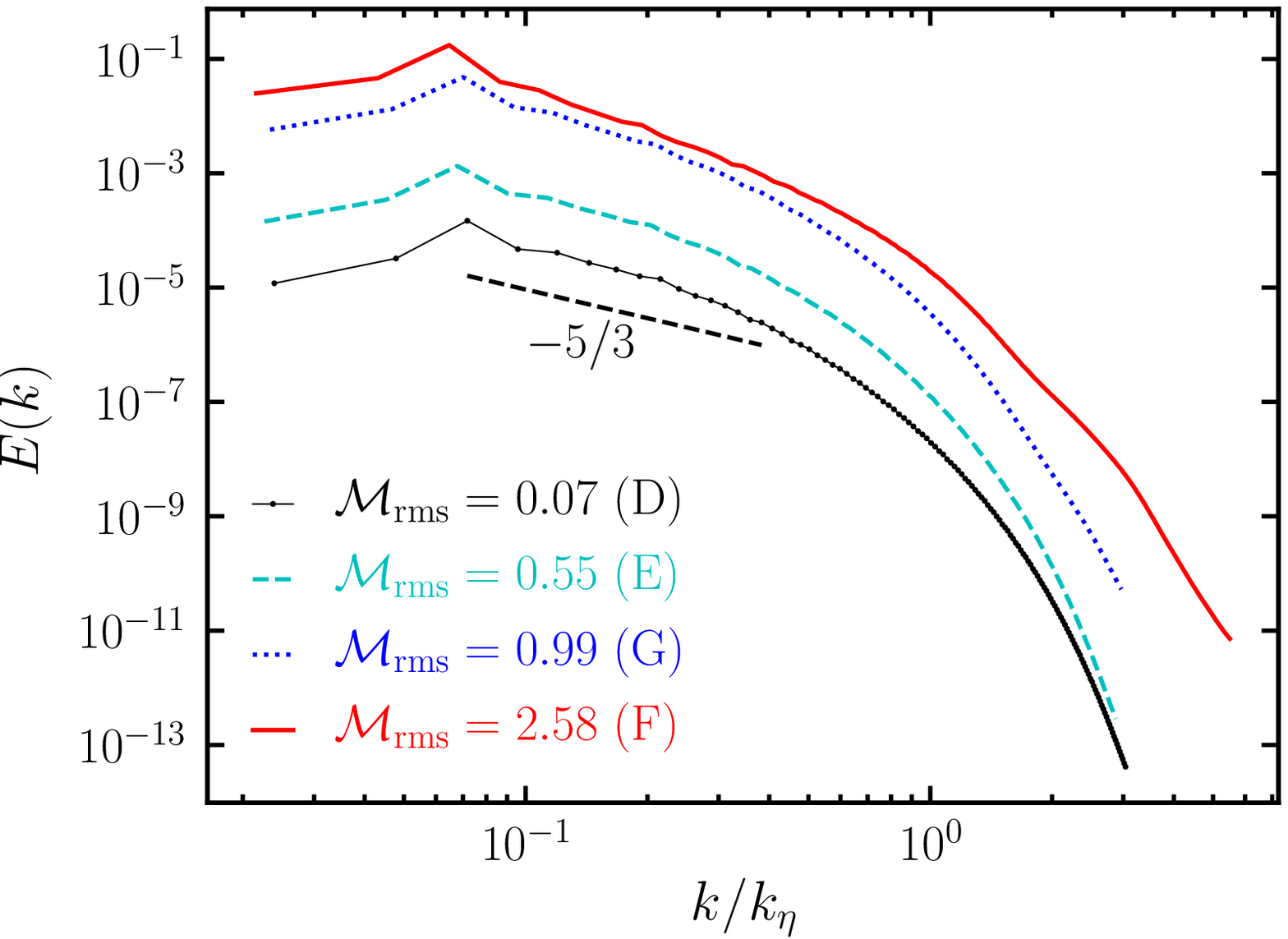}}
\caption{Power spectra for simulations of different
${\cal M}_{\rm rms}$: ${\cal M}_{\rm rms}=0.07$ (solid dotted black curve),
$0.55$ (dashed cyan curve), $0.99$ (dotted blue curve), and $2.58$ (red curve).
The dashed black curve shows the Kolmogorov $-5/3$ law.}
\label{ppower_M}
\end{figure}

To investigate how coagulation depends on the Mach number $\mathcal{M}_{\rm rms}$, we performed simulations for different $\mathcal{M}_{\rm rms}$ ranging from 0.07 to 2.58 while keeping ${\rm Re}$ and $\langle\bar{\epsilon}\rangle$ fixed
(see the details of the simulation setup in \Tab{tab:helical}).
As shown in \Fig{ppower_M}, the power spectra
follow the classical Kolmogorov $-5/3$ law.
Because the Reynolds number that can be achieved in DNS studies
is much lower than the one in the ISM,
large-scale separation of turbulence is not observed.

\begin{figure}
\resizebox{\hsize}{!}{
\includegraphics{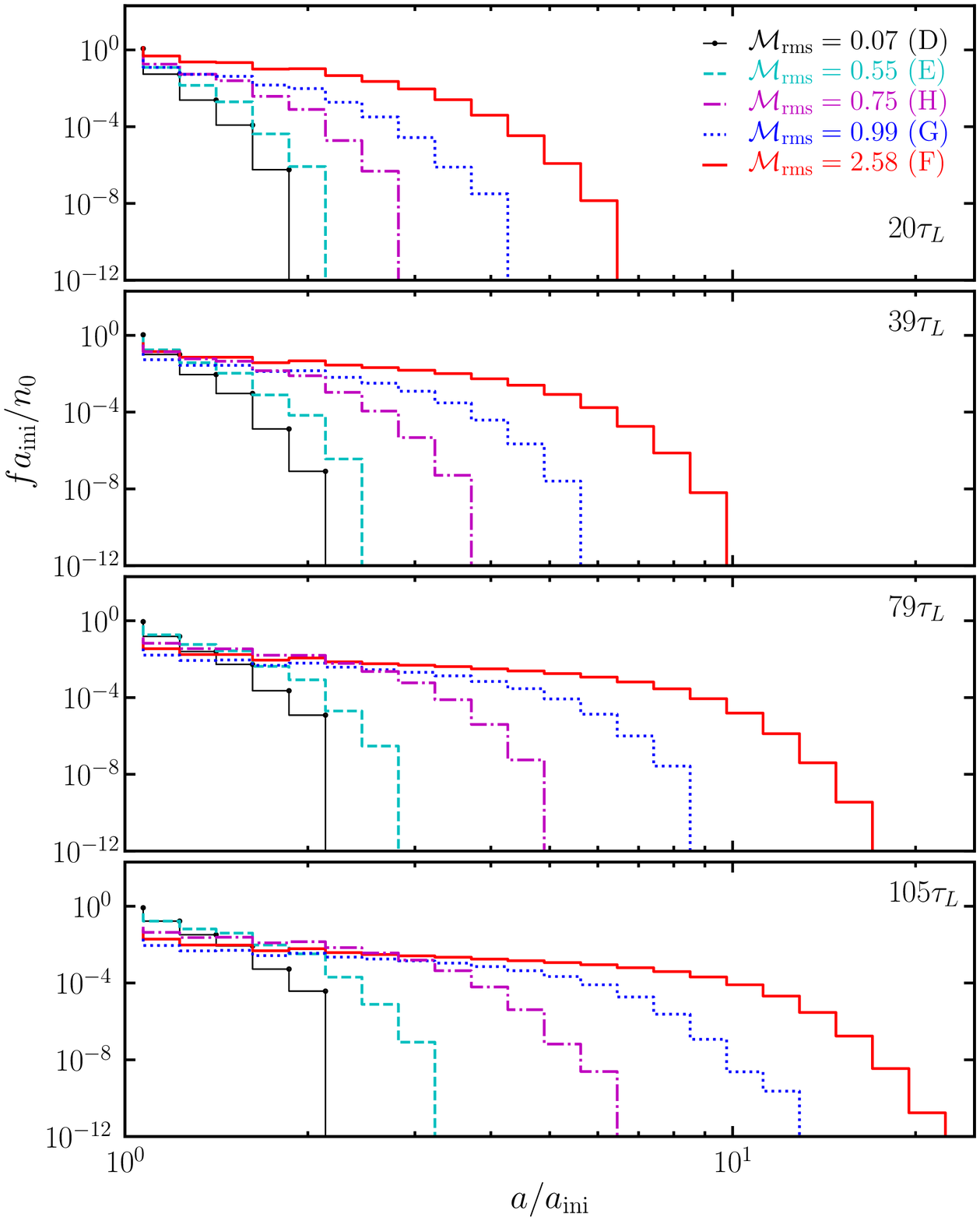}}
\caption{Time evolution of $f(a,t)$
for simulations in \Fig{ppower_M} and for
an additional simulation run H in \Tab{tab:helical}.}
\label{f_cond0_coa}
\end{figure}

Next, we inspected the time evolution of the dust size distribution $f(a,t)$. As shown in \Fig{f_cond0_coa}, the tail of $f(a,t)$ widens with increasing ${\cal M}_{\rm rms}$. The broadening of $f(a,t)$ is slowest for the nearly incompressible flow with ${\cal M}_{\rm rms}=0.07$ (solid dotted black curve). A transition is observed when the flow pattern changes from subsonic (${\cal M}_{\rm rms}\sim 0.5$) to transonic or supersonic (${\cal M}_{\rm rms}\gtrsim 1$)\footnote{A turbulent flow may be categorised in the following way according to Mach number: subsonic range (${\cal M}_{\rm rms}<0.8$), transonic range ($0.8< {\cal M}_{\rm rms}<1.3$), and the supersonic range ($1.3<{\cal M}_{\rm rms}<5.0$).}, where the broadening and the extension of the tail of $f(a,t)$ become prominent. This is further evidenced by the simulations with ${\cal M}_{\rm rms}=0.75$ (dash dotted magenta curve) and ${\cal M}_{\rm rms}=0.99$ (dotted blue curve), in the intermediate transonic regime. The supersonic case with ${\cal M}_{\rm rms}=2.58$ displays a
significant broadening of the tail.

\begin{figure}
\resizebox{\hsize}{!}{
\includegraphics{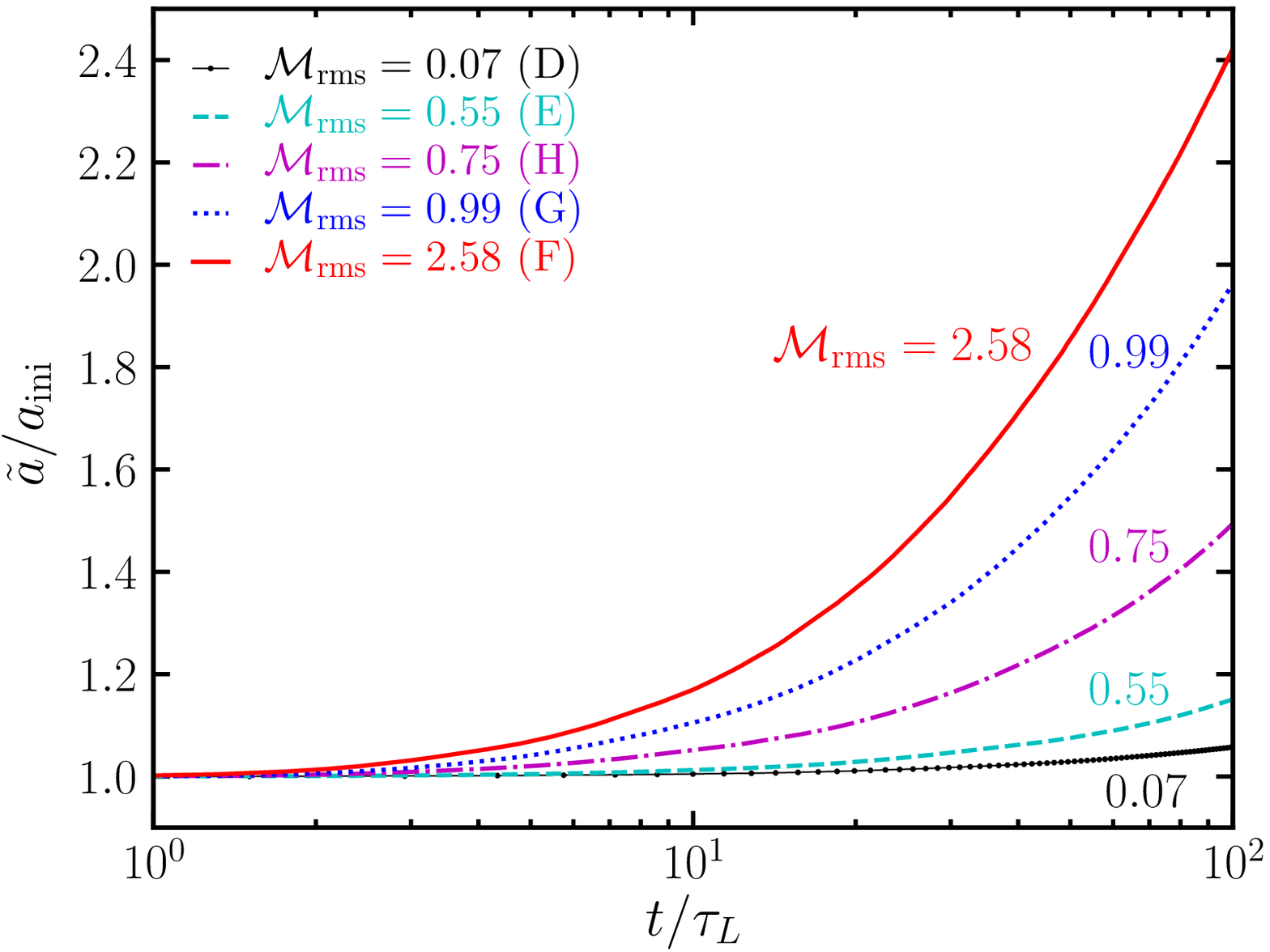}}
\caption{Time evolution of $\tilde{a}/a_{\rm ini}$
for simulations in \Fig{f_cond0_coa}.}
\label{a1_comp_helical}
\end{figure}

\Fig{a1_comp_helical} shows the time evolution of the mean radius $\tilde{a}$ normalised by the initial size of particles. It is obvious that $\tilde{a}/a_{\rm ini}$ increases with increasing ${\cal M}_{\rm rms}$. Although this does not say much about tail effects, $\tilde{a}$ is a good measurement of the mean evolution of $f(a,t)$.

\begin{figure}
\resizebox{\hsize}{!}{
\includegraphics{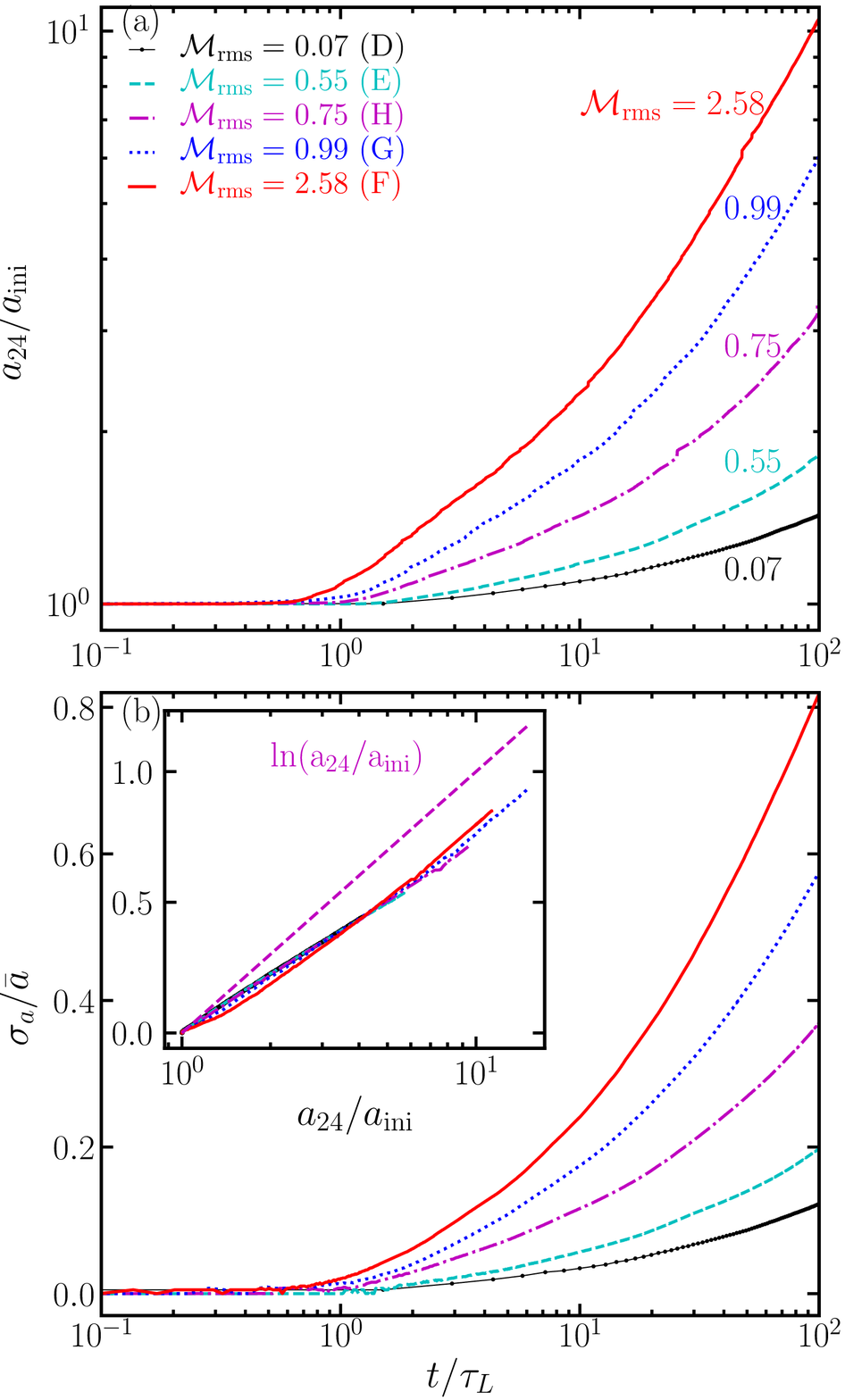}}
\caption{Time evolution of (a) the $a_{24}$ and (b) $\sigma_a/\tilde{a}$
for simulations in \Fig{f_cond0_coa}. The inset shares the same
y-axis as the main plot in panel (b). The dashed magenta curve
in the inset shows ${\rm ln}(a_{24}/a_{\rm ini})$.}
\label{a24_comp_helical}
\end{figure}

According to \Eq{kernel_gen}, the coagulation rate depends on the total cross section of the two colliding particles. Therefore growth by coagulation is sensitive to the large tail of $f(a,t)$. As discussed in Section~\ref{sec:dia}, the tail of $f(a,t)$ can be characterised by $a_{24}$, that is, the 24th normalised moment. \Fig{a24_comp_helical}(a) shows that the rate of increase of $a_{24}$ increases with ${\cal M}_{\rm rms}$.
The corresponding relative dispersion of $f(a,t)$, $\sigma_a/\tilde{a}$, is shown in \Fig{a24_comp_helical}(b), which exhibits the same ${\cal M}_{\rm rms}$ dependence as $a_{24}$. However,  the form of $\sigma_a/\tilde{a}$ as a function of $a_{24}/a_{\rm ini}$ is essentially independent of ${\cal M}_{\rm rms}$ , as shown by the inset in \Fig{a24_comp_helical}(b).

\begin{figure*}
\resizebox{\hsize}{!}{
\includegraphics{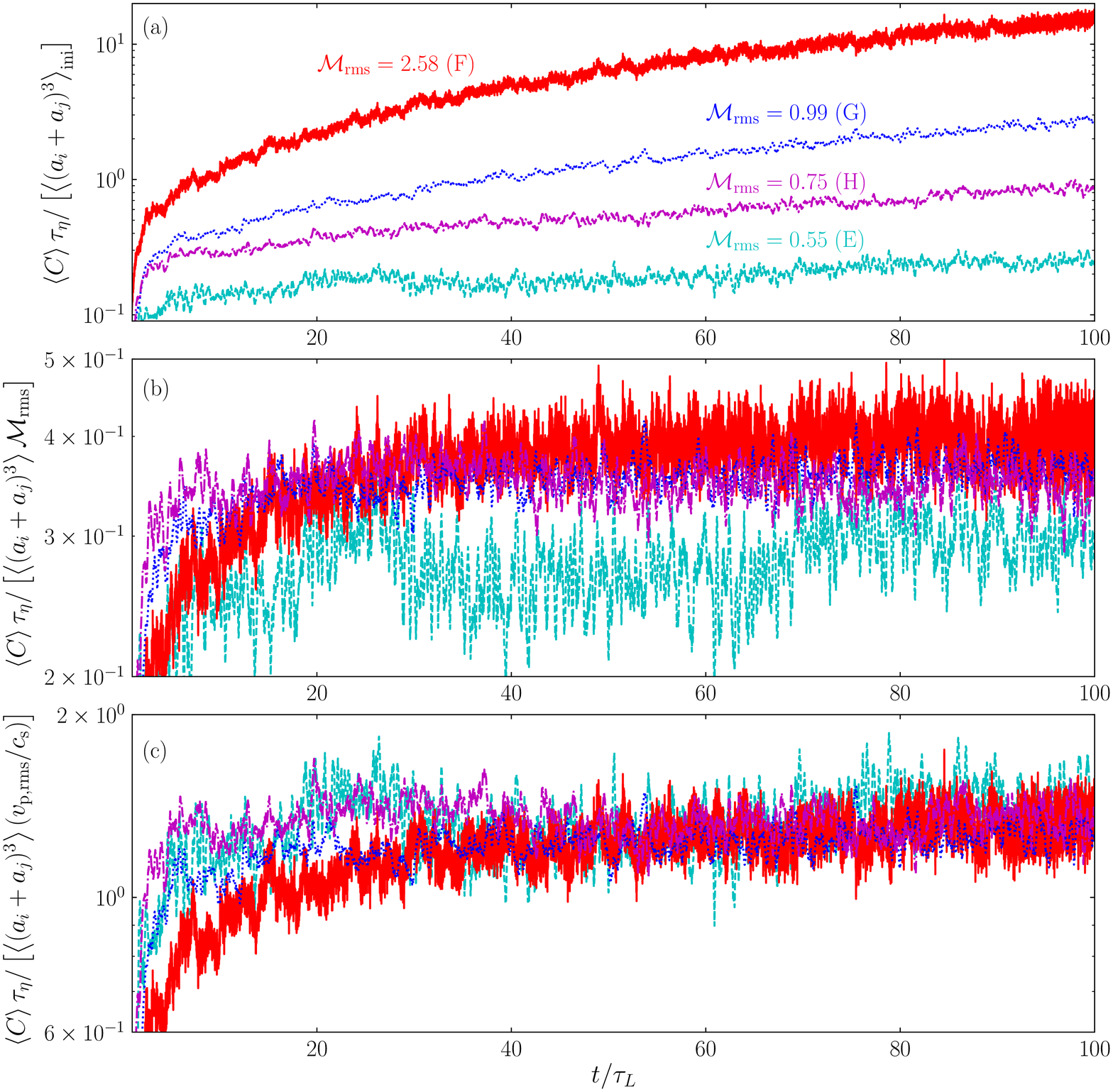}}
\caption{Measured collision kernel $\langle C \rangle$ normalised to $(a_i+a_j)^3$. Panel (a) shows the case of constant (initial) $a_i$, while panel (b) shows the case where $(a_i+a_j)^3$ evolves and where ${\cal M}_{\rm rms}$ is also included in the normalisation. The simulations are the same as in \Fig{f_cond0_coa}.
  Panel (c) shows the same normalisation as panel (b), but with ${\cal M}_{\rm p,rms}$.}
\label{NCOAGPM}
\end{figure*}

As mentioned in section~\ref{sec:theory}, the mean collision kernel $\langle C_{ij}\rangle$ depends on ${\cal M}_{\rm rms}$. \Fig{NCOAGPM}(a) shows the collision kernel $\langle C_{ij}\rangle$ normalised according to $(a_i+a_j)_{\rm ini}^3$, that is, the initial particle size. The Saffman-Turner model should not apply to coagulation of inertial particles in compressible turbulence as it assumes that particles act as passive tracers and are advected by the turbulent motion of the carrier. In spite of this, $\langle C_{ij}\rangle$ appears to scale with particle size as $a^3$, which is shown in \Fig{NCOAGPM}(b), where $\langle C_{ij}\rangle$ is normalised to $\mathcal{M}_{\rm rms}\,(a_i+a_j)^3$.
The reason for this is not obvious.
We recall, however, that we consider turbulence in highly compressible flows, and
more importantly, that the trajectories of inertial particles tend to deviate from the flow.
This leads to higher particle densities by compaction and clustering
in the convergence zones in between vortex tubes \citep{Maxey87}.
Moreover, depending on the particle masses, it may also lead to the formation of caustics,
which are the singularities in phase-space of suspended inertial particles
\citep{falkovich2002acceleration, wilkinson2005caustics}.
This will lead to large velocity differences between colliding particles and thus to large $\langle C_{ij}\rangle$.
The net result is a rather complex coagulation process, where $\langle C_{ij}\rangle$ varies strongly from one location to the next,
which is further discussed below.

\begin{figure}
\resizebox{\hsize}{!}{
\includegraphics{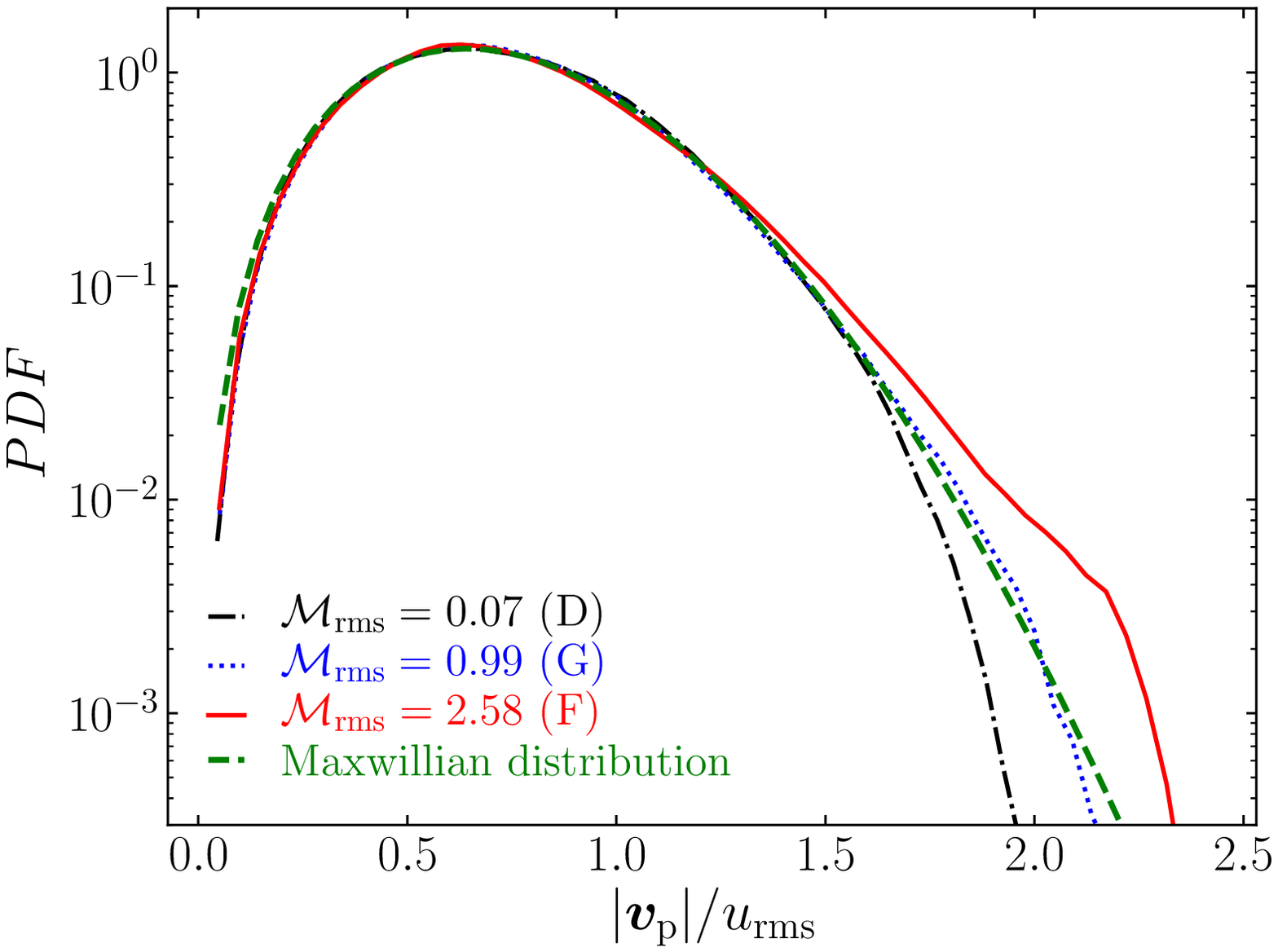}}
\caption{Particle velocity distribution at $105\tau_L$. The dashed line
is a Maxwellian fit of the particle velocity of run G.}
\label{vp_maxwellian_SW512condens0_coag_nu2p5em3_L3_f3_cs0p5_nuS8_256CPU}
\end{figure}

According to \Fig{NCOAGPM}(a), $\langle C_{ij}\rangle$ exhibits a clear increase from subsonic to supersonic turbulence
(cf. ${\cal M}_{\rm rms}=0.55$, cyan curve in \Fig{NCOAGPM}(a) and ${\cal M}_{\rm rms}=2.58$, red curve in \Fig{NCOAGPM}(a)).
As we argued in section~\ref{sec:theory}, $\langle C_{ij}\rangle$ is proportional to ${\cal M}_{\rm rms}$
under the assumption of Maxwellian velocity distributions. \Fig{NCOAGPM} (b) shows $\langle C_{ij}\rangle$
normalised also by ${\cal M}_{\rm rms}$.
We note that a linear scaling seems to be applicable from the subsonic regime to the supersonic regime.
This means that the simple analytical theory is reasonable.
The cyan curve deviates from other curves because
the initial Stokes number for this simulation is about 10, as listed in \Tab{tab:helical}. The Stokes number dependence
of $\langle C_{ij}\rangle$ can be further confirmed by
\Fig{NCOAGPM}(c).
As $\langle C_{ij}\rangle$ is determined by the relative velocity
of colliding pairs, we normalised $\langle C_{ij}\rangle$ by
$v_{\rm p,rms}/c_{\rm s}$
, as shown in \Fig{NCOAGPM}(c). It is obvious that $\langle C_{ij}\rangle$ scales linearly with $v_{\rm p,rms}/c_{\rm s}$ up to the supersonic
regime.
\Fig{vp_maxwellian_SW512condens0_coag_nu2p5em3_L3_f3_cs0p5_nuS8_256CPU} shows the distribution of the magnitudes
of the particle velocities, which indeed is very similar to a Maxwellian velocity distribution.
It also shows that particle velocities become higher with increasing ${\cal M}_{\rm rms}$. Especially the tail of the particle velocity distribution becomes more populated. This indicates that stronger shocks accelerate inertial particles more and may therefore increase the coagulation rate.

According to \Eq{kernel_gen}, the collision kernel is determined by the relative velocity $\bm{w}_{ij}$
and the relative separation $\Delta \bm{r}$ of two colliding particles.
The former scenario is known as caustics \citep{Wilkinson06} and the latter as clustering \citep{GM16}, as discussed above.
Our simulations involve coagulation, which leads to evolution of $\alpha(a,t)$. This makes it difficult to analyse clustering and caustics based on these simulations.
Below we try to understand the ${\cal M}_{\rm rms}$-dependence based on the spatial distribution and velocity statistics of the particles.

\begin{figure*}\begin{center}
\includegraphics[width=\textwidth]{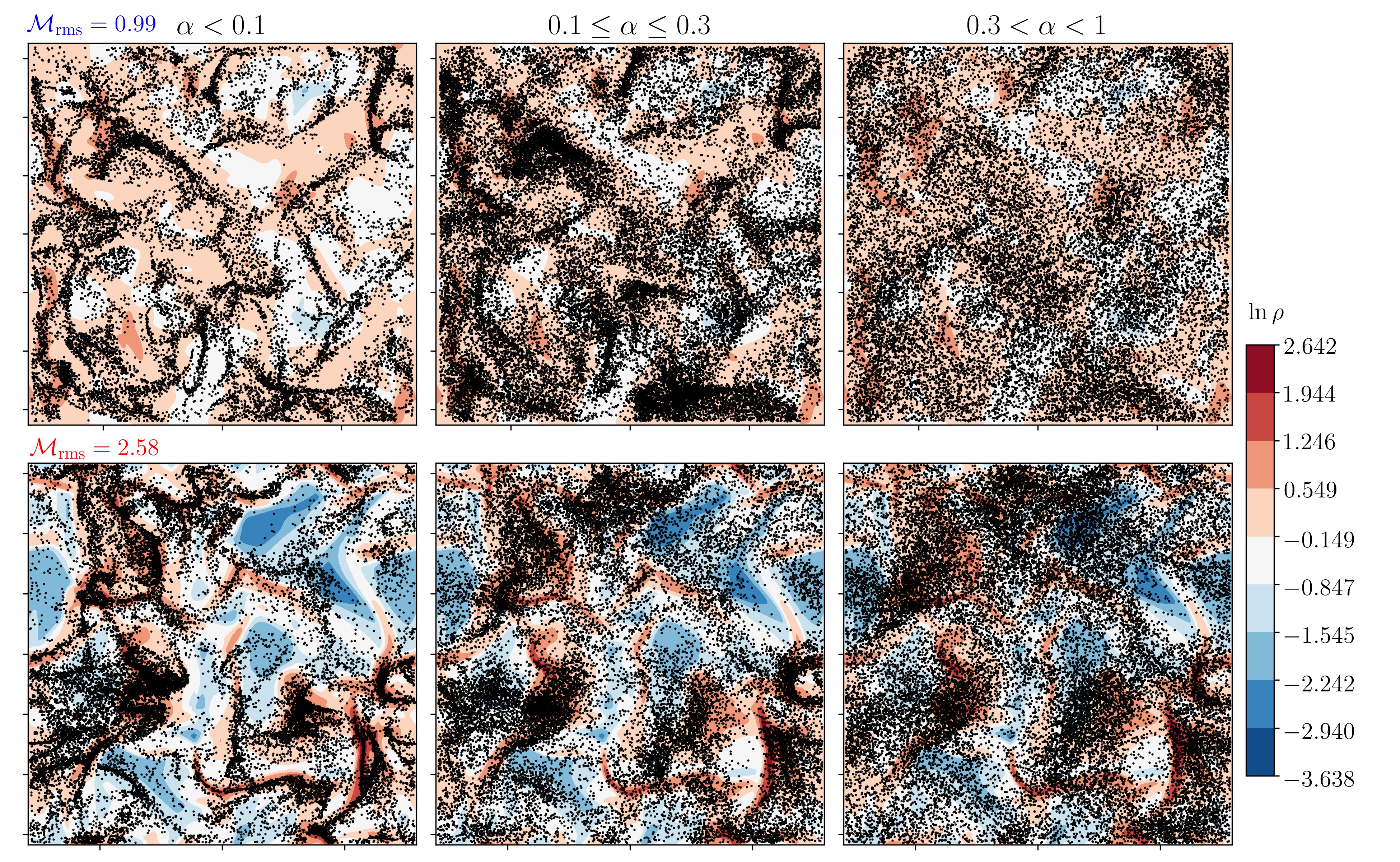}
\end{center}
\caption{Spatial distribution of inertial particles and the gas density at 80 time units
in a slab with thickness $\eta$ for run F.}
\label{alpha_slice}
\end{figure*}

\begin{figure*}\begin{center}
\includegraphics[width=\textwidth]{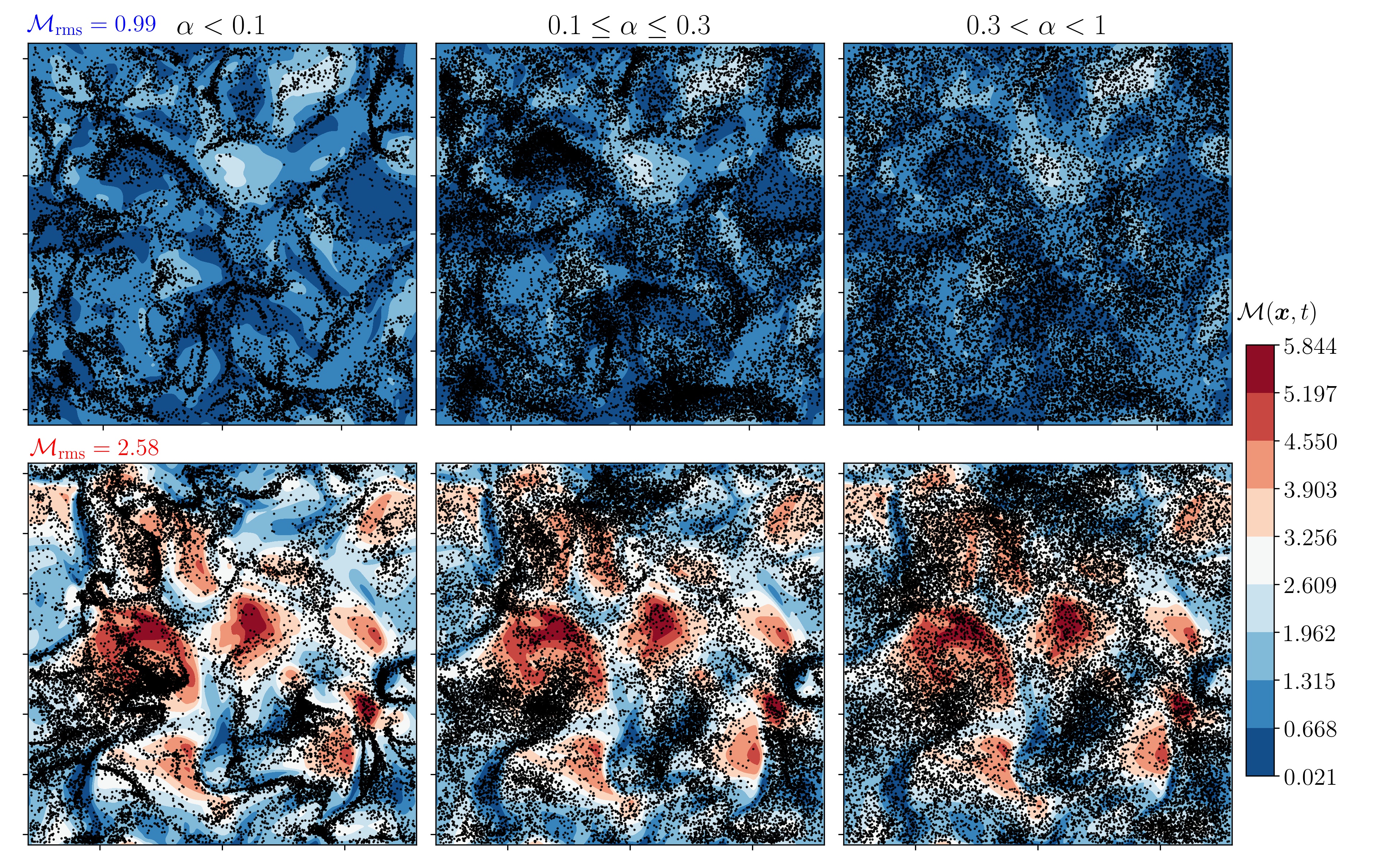}
\end{center}
\caption{Same as \Fig{alpha_slice}, but with ${\cal M}({\bm x},t)$ as
the contour map.}
\label{Ma_particles_slice}
\end{figure*}

As shown in \Fig{alpha_slice}, the spatial distribution of particles exhibits different behaviours in the three $\alpha$ ranges we considered. When $\alpha<0.1$, particles tend to be trapped in regions where high gas density occurs. This is consistent with the findings of \citet{Hopkins16} and \citet{Lars19_Small}, even though coagulation was not considered in their studies. When $0.1\le\alpha\le0.3$, particles still accumulate in the high-density regions, but are also spread out in regions with low gas density. This dispersion is expected as $\tau_i$ increases. Finally, when $\alpha>0.3$, particles more or less decouple from the flow, demonstrating essentially a random-walk behaviour.
When we compare with ${\cal M}(\bm{x},t)$ instead of $\ln{\rho}$, we see that particles accumulate in regions with low local ${\cal M}(\bm{x},t)$, as shown in \Fig{Ma_particles_slice}. That is, low ${\cal M}(\bm{x},t)$
corresponds to high $\ln{\rho}(\bm{x},t)$.
The physical picture is the following. Strong shocks generated in these local supersonic regions push particles to low ${\cal M}(\bm{x},t)$ regions, which is then how particle densities increase due to compression of the gas.
This {\it \textup{compaction}} of particles is different from the {\it \textup{fractal clustering}} of inertial particles, which mainly occurs as a result of accumulation of particles in the convergence zones between vortices.
Statistically, the spatial distribution of particles can be characterised by $g(r)$, which contributes to the mean collision rate as expressed in \Eq{kernel_gen}.
However, $g(r)$ is only useful as a diagnostic for a mono-dispersed particle distribution or fixed size bins \citep{pan2011turbulent}. Therefore we only show the spatial distributions of particles and do not go into details about the quantitative statistics.

\begin{figure}\begin{center}
\includegraphics[width=0.5\textwidth]{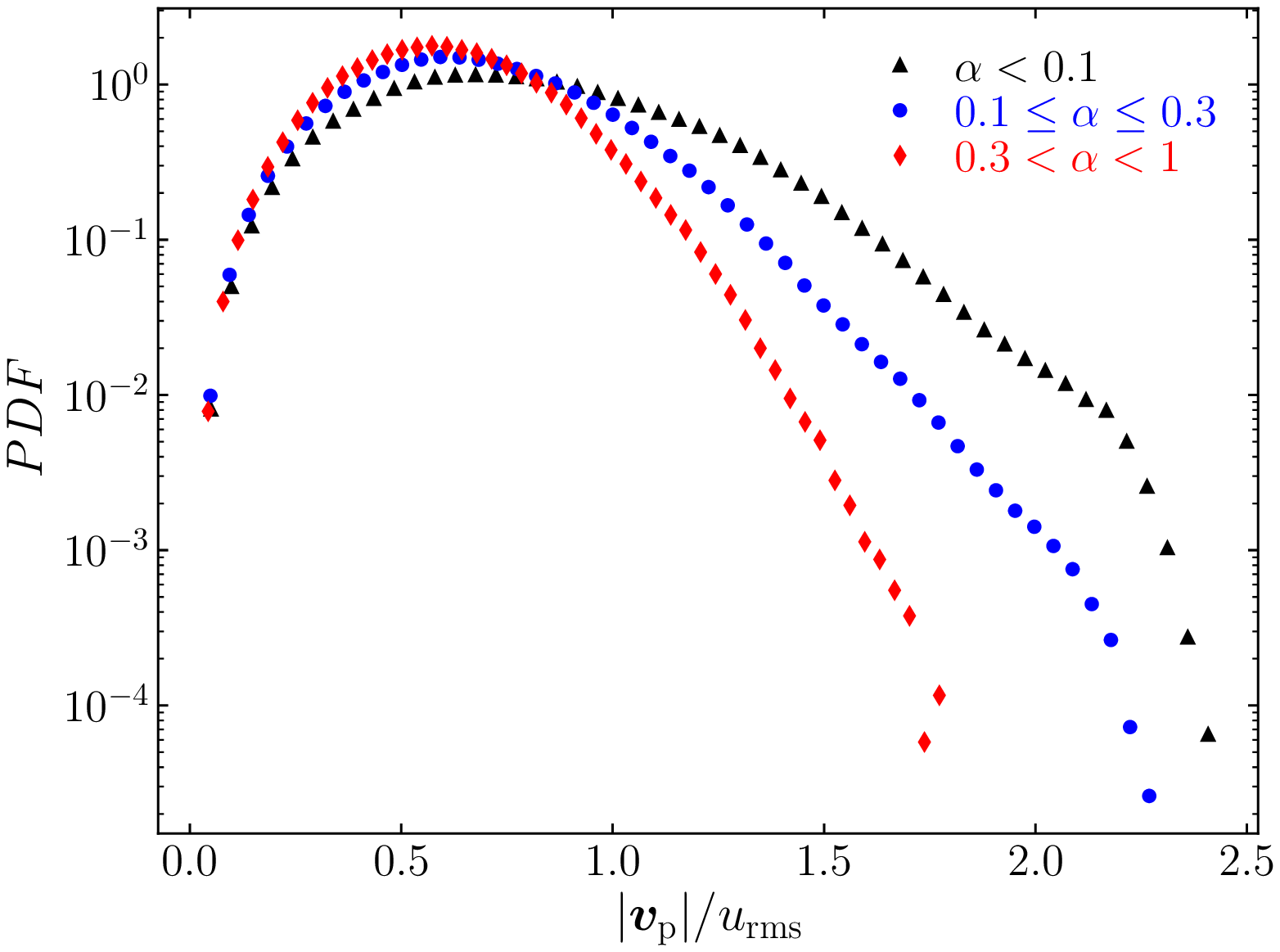}
\end{center}
\caption{Corresponding PDF for
\Fig{alpha_slice}.}
\label{vp_pdf}
\end{figure}

As the collision kernel $\langle C_{ij} \rangle$ is also
determined by $|{\bm w}_{ij}|$, we also examined the magnitude
of particle velocities for different ranges of $\alpha$.
\Fig{vp_pdf} shows the PDF of $|\bm{v}_p|/u_{\rm rms}$ for
$\alpha<0.1$, $0.1\le\alpha\le0.3$, and $\alpha>0.3$, respectively.
It is evident that the velocity magnitude of particles
that are coupled to the flow is higher than that of particles that are decoupled from the flow.
Thus, the ${\cal M}_{\rm rms}$ dependence of $\langle C_{ij} \rangle$ could very well
be due to enhanced caustics and compression-induced concentration with increasing ${\cal M}_{\rm rms}$.

\begin{figure}
\resizebox{\hsize}{!}{
\includegraphics{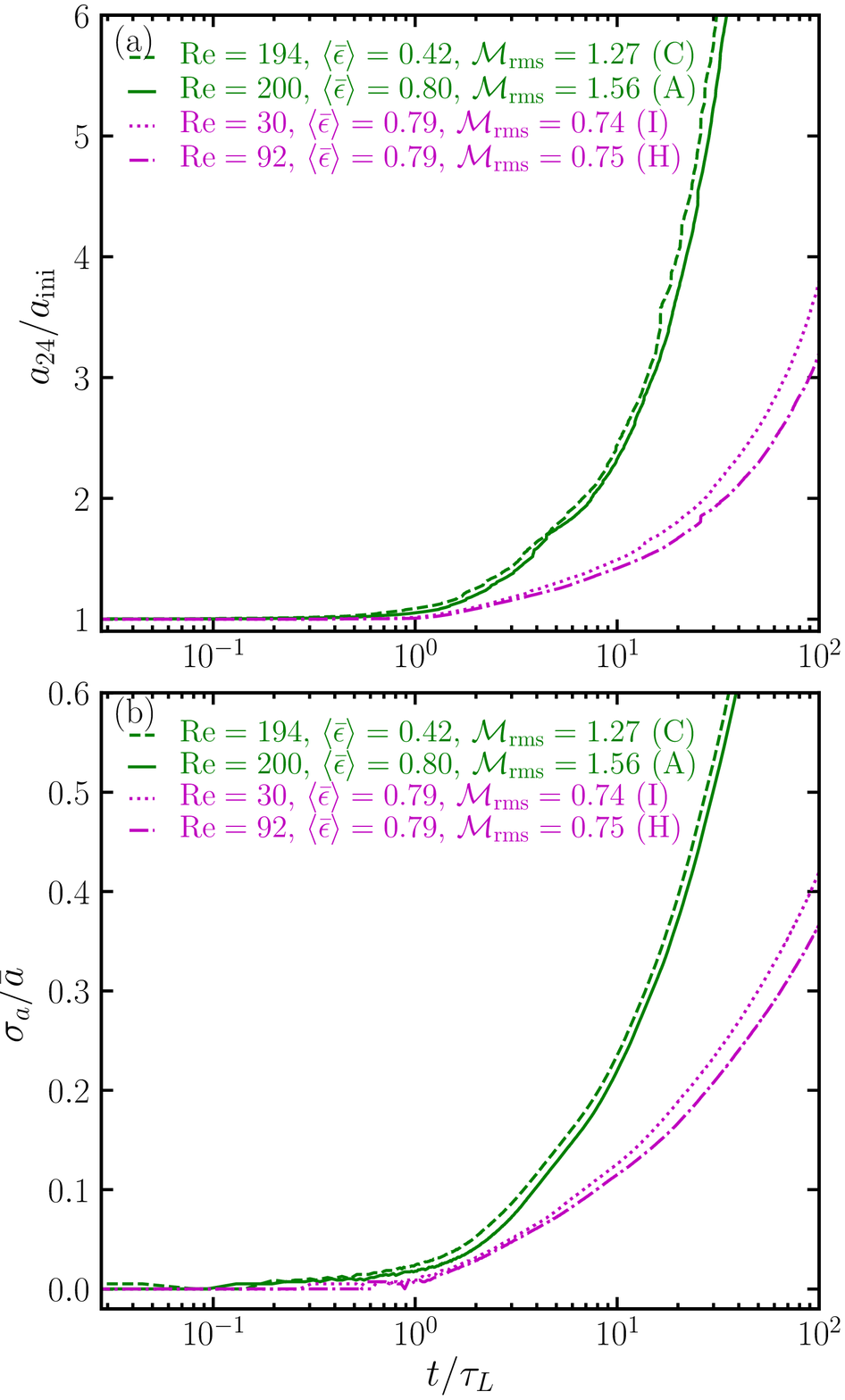}}
\caption{Time evolution of (a) $a_{24}$ and (b) dispersion of $f(a,t)$
for different $\langle\bar{\epsilon}\rangle$ and ${\rm Re}$ with the same
${\cal M}_{\rm rms}$ (see runs A, C, H, and I in \Tab{tab:helical}).}
\label{a24_comp_Re_epsi}
\end{figure}

In incompressible turbulence, the collision kernel depends on $\tau_\eta$, which is determined by
$\langle\bar{\epsilon}\rangle$,
while it is insensitive to ${\rm Re}$ \citep{Grabowski_2013, li2017effect}.
We examined the $\langle\bar{\epsilon}\rangle$ and ${\rm Re}$
dependences of $a_{24}$ and $\sigma_a$ in compressible turbulence.
As shown in \Fig{a24_comp_Re_epsi}, $a_{24}$ and $\sigma_a/\tilde{a}$
have only a weak dependence on $\langle\bar{\epsilon}\rangle$ in the supersonic
regime
(e.g. simulations A and C have similar ${\cal M}_{\rm rms}$ but differ
by a factor of two in $\langle\bar{\epsilon}\rangle$).
By inspection of \Fig{a24_comp_Re_epsi}, changing ${\rm Re}$
(run H and I)
does not obviously affect $a_{24}$ and $\sigma_a/\tilde{a}$ 
in the transonic regime,
which may seem to be consistent with the simulation results for
incompressible turbulence.

\section{Discussion}

We have observed that as ${\cal M}_{\rm rms}$ increases, the tail of the distribution increases. This poses mainly two questions: (1) how the compression-induced density variation (simply compaction) affects
$\langle R_{ij} \rangle$, and (2) how the velocity dispersion of particles due to
shocks affects $\langle C_{ij} \rangle$.

According to \Eq{kernel_gen},
the coagulation rate $\langle R_{ij}\rangle$ is determined
by $g(a,r)$ and $|\bm{w}_{ij}|,$ as discussed in section~\ref{sec:results}.
Particles tend to stay in regions where the gas density is high due to
shock-induced compaction. With larger ${\cal M}_{\rm rms}$, the gas
density fluctuations become stronger. This leads to somewhat higher
concentrations of particles, especially for particles with $\alpha\le 0.3$.
It is important to note that particles accumulate in low ${\cal M}\left(\bm{x},t\right)$
regions, where the gas density is high because particles are pushed
to lower ${\cal M}\left(\bm{x},t\right)$ regions by the shocks.
Supersonic flows cover a wide range of ${\cal M}\left(\bm{x},t\right)$,
which results in stronger density variations of particles.
Local concentrations of particles might lead
to higher $\langle R_{ij} \rangle$.
Therefore the enhanced $\langle R_{ij} \rangle$ with increasing
${\cal M}_{\rm rms}$ is indeed due to
the change in the flow structure.
In addition to the compressibility,
fractal clustering due to inertia effect
(large St) of particles
might also enhance $\langle C_{ij}\rangle$.

Higher ${\cal M}_{\rm rms}$ results in stronger shocks and thus higher particle velocities, which also leads to larger $\langle C_{ij} \rangle$.
In particular, for supersonic flows, the local fluctuations of ${\cal M}\left(\bm{x},t\right)$ are strong.
This leads to significant local dispersion in the particle velocities and a consequent enhancement of $\langle C_{ij} \rangle$.
The coagulation rate time series almost collapse on top of each
other when normalised by $v_{\rm p,rms}/c_{\rm s}$ up to the supersonic
regime. This indicates that the simple description of the collision kernel, \Eq{eq:cij}, applies up to the supersonic regime.
As particles grow larger in the simulation, they decouple 
from the flow. Statistically, this decoupling can be roughly described by
the difference between $u_{\rm rms}$ and $v_{\rm p,rms}$.
This probably causes these curves to collapse on each other
when $\langle C_{ij} \rangle$ is normalised by $v_{\rm p,rms}/c_{\rm s}$.
We therefore propose
that $\langle C_{ij} \rangle$ is proportional to $v_{\rm p,rms}/c_{\rm s}$
instead of ${\cal M}_{\rm rms}$ in \Eq{eq:cij}.

Since the inertial range is determined by ${\rm Re}$, we also examined
how $\langle C_{ij} \rangle$ depends on ${\rm Re}$.
As shown in \Fig{a24_comp_Re_epsi}, the  $\rm{Re}$ dependence is weak because $\tau_i \ll \tau_L$. \citet{pan2014turbulence}
have suggested that the collision kernel is independent of $\rm{Re}$
in the subsonic regime. We show here that this also appears to apply to the transonic regime and likely also the supersonic regime.
As we discussed in section~\ref{sec:flow}, for incompressible turbulence, $\langle C_{ij} \rangle$ is determined by $\langle\bar{\epsilon}\rangle$ through $\tau_\eta$.
\Fig{a24_comp_Re_epsi} shows that the $\langle\bar{\epsilon}\rangle$ dependence observed in incompressible flows vanishes in compressible flows, however, which is quite expected. We demonstrated that the coagulation rate of inertial particles is mainly affected by ${\cal M}_{\rm rms}$, essentially the level of compression of the flow. 
We conclude that ${\cal M}_{\rm rms}$ is the main parameter determining $\langle C_{ij} \rangle$ in the trans- and supersonic regimes. 

The pioneering work of \citet{1955_Saffman} suggested that
$\langle C_i\rangle\sim \tau_\eta^{-1}$ for
specified sizes of particles. Because $\tau_\eta=(\nu/\langle\bar{\epsilon}\rangle)^{1/2}$,
$\langle C_i\rangle\sim \langle\bar{\epsilon}\rangle^{1/2}$, which has been confirmed in many studies of incompressible turbulence.
More importantly, all simulation works have found that $\langle C_i\rangle$
is independent of Re \citep[see][and the references therein]{Grabowski_2013, li2017effect}. This is contradictory, however, because
$\langle\bar{\epsilon}\rangle$ is a parameter that is determined by Re.
We argue that this is because of the forcing term in the N-S equation,
which is inevitable in turbulence simulations. This forcing term
invokes a third dimensional parameter $\langle\bar{\epsilon}\rangle$
that determines $\langle C_i\rangle$ in incompressible turbulence.
For compressible turbulence, however, our study showed
that $\langle C_i\rangle$ is determined by ${\cal M}_{\rm rms}$ alone
and independent of $\langle\bar{\epsilon}\rangle$.
We can then avoid the contradiction described above. However, this contradiction
is a more fundamental problem that requires a solution,
but this is beyond the scope of our study.
In short, compressible turbulence implies that $\langle\bar{\epsilon}\rangle$ is not only determined by
Re, but also by ${\cal M}_{\rm rms}$.
Even though $\langle\bar{\epsilon}\rangle$ does not affect $\langle C_i\rangle$ directly,
it affects the Stokes number ${\rm St}=\tau_{\rm i}/\tau_\eta$.
Therefore the $\langle C_i\rangle$ time series collapse on top of each other
when they are normalised by $\tau_\eta$ in \Fig{NCOAGPM}(c).

\citet{andersson2007advective}
proposed that in 2D compressible flow
with a Gaussian random velocity, fractal particle (inertialess) clustering
can lead to a higher coagulation rate. In addition to the
aforementioned assumptions, the collision rate suggested in
\citet{andersson2007advective} involves the fractal dimension $D_2$,
which is difficult to measure in our case because
we have an evolving size distribution of particles. A direct comparison between our simulation and the theory of \citet{andersson2007advective} is therefore not feasible. Nevertheless, we do
observe fractal clustering in our simulation, which could
indeed elevate the coagulation rate because
\Fig{vp_maxwellian_SW512condens0_coag_nu2p5em3_L3_f3_cs0p5_nuS8_256CPU}
shows that high ${\cal M}_{\rm rms}$ flow results in higher particle velocities.
Because we have a wide range of Stokes numbers, the fractal
clustering \citep{falkovich2001particles} could also be enhanced.

\section{Summary and conclusion}
Coagulation of inertial particles in compressible turbulence was investigated by direct and shock-capturing numerical simulations. 
Particle interaction was tracked dynamically in a Lagrangian manner, and the consequential coagulation events were counted at each simulation time step.
We specifically explored the Mach-number dependence of the coagulation rate and the effects on the widening of the particle-size distribution.
To our knowledge, this is the first time that this has been done.

We showed that the coagulation rate is determined by Mach number
${\cal M}_{\rm rms}$ in compressible turbulence.
This is fundamentally different from the incompressible case,
where the coagulation rate is mainly determined by $\langle\bar{\epsilon}\rangle$
through the Kolmogorov timescale.

The dispersion or variance of $f(a,t)$, $\sigma_a$, increases with increasing ${\cal M}_{\rm rms}$. We also note that $\sigma_a$ is a simple and more or less universal function of the size of the largest particles, measured by $a_{24}$, and is apparently independent of ${\cal M}_{\rm rms}$. 

All effects on coagulation increase progressively with ${\cal M}_{\rm rms}$, which shows the importance of  compressibility for coagulation processes. Taken at face value, our simulations appear to suggest that existing theories of the ${\cal M}_{\rm rms}$ dependence of $\langle C_{ij}\rangle$, which imply an underlying linear scaling with ${\cal M}_{\rm rms}$ , are correct to first order, but we cannot draw any firm conclusions at this point. For this we will need more simulations with a wider range of ${\cal M}_{\rm rms}$ values.
We note that $\langle C_{ij}\rangle$ scales as
$\langle C_{ij}\rangle \sim (a_i + a_j)^3\,\mathcal{M}_{\rm rms}/\tau_\eta$.
When the collision kernel $\langle C_{ij}\rangle$ is normalised
by $v_{\rm p,rms}/c_{\rm s}$,
the curves collapse on top of each other. We therefore suggest that
$\langle C_{ij}\rangle$ is proportional to
$v_{\rm p,rms}/c_{\rm s}$, rather than $\mathcal{M}_{\rm rms}$.
This finding may serve as a benchmark for future studies of coagulation of dust grains in highly compressible
turbulence.

We propose two mechanisms that might be behind the ${\cal M}_{\rm rms}$ dependence of the broadening of $f(a,t)$
even though it is still not fully understood due to the non-equilibrium nature of the coagulation process
in compressible turbulence. The first mechanism is the compaction-induced concentration of particles.
Supersonic flow exhibits stronger fluctuations of local ${\cal M}(\bm{x},t)$.
The consequent vigorous shocks compact small particles (e.g. $\alpha<0.3$)
into low-${\cal M}(\bm{x},t)$ regions. This leads to high densities of particles ($n_i$)
and then potentially
to a higher coagulation rate $\langle R_{ij}\rangle$.
The second mechanism is larger dispersion of particle velocities caused by stronger shocks.
Again, stronger local fluctuations ${\cal M}(\bm{x},t)$ lead to a larger dispersion of particle velocities,
which increases the coagulation rate.

Simulating the coagulation problem in compressible and supersonic turbulence, we achieved ${\cal M}_{\rm rms}=2.58$, but with a non-astrophysical scaling. This is smaller than the ${\cal M}_{\rm rms}\ge 10$ observed in cold clouds. To explore whether  a saturation limit of the ${\cal M}_{\rm rms}$ dependence of the coagulation rate exists, a direct numerical simulation coupled with coagulation would have to reach at least ${\cal M}_{\rm rms}\sim 10$. 

We also note that the simulated systems in our study have flow timescales (turn-over times) that are of the same order as the coagulation timescale, that is,  $\tau_{\rm c}/\tau_L< 1$, which is computationally convenient, but very different from dust in the ISM, for example, where $\tau_c/\tau_L\gg1$. Nonetheless, our study provides a benchmark for simulations of dust-grain growth by coagulation in the ISM and other dilute astrophysical environments.

Reaching really high ${\cal M}_{\rm rms}$, and astrophysical scales in general is currently being explored. Fragmentation is also omitted in this study, which may overestimate the coagulation rate.
Adding fragmentation is a topic for future work. 

\section*{Acknowledgement}

Xiang-Yu Li wishes to thank Axel Brandenburg, Nils Haugen, and Anders Johansen
for illuminating discussions about the simulation code used in this study,
the {\sc Pencil Code}. Lars Mattsson wishes to thank the Swedish Research Council
(Vetenskapsrdet, grant no. 2015-04505) for financial support.
We thank the referee for very helpful suggestions to improve
the manuscript.
Our simulations were performed using resources provided by the Swedish National
Infrastructure for Computing (SNIC) at the Royal Institute of Technology in Stockholm,
Link\"oping University in Linköping, and Chalmers Centre for Computational Science
and Engineering (C3SE) in Gothenburg. The {\sc Pencil Code} is freely
available on \url{https://github.com/pencil-code/}.
The authors also thank the anonymous reviewer for constructive
comments on the paper.

%\bibliographystyle{aa}
%\bibliography{database}

\begin{thebibliography}{70}
\expandafter\ifx\csname natexlab\endcsname\relax\def\natexlab#1{#1}\fi

\bibitem[{Abrahamson(1975)}]{Abrahamson75}
Abrahamson, J. 1975, Chemical Engineering Science, 30, 1371

\bibitem[{Aldous(1999)}]{Aldous99}
Aldous, D.~J. 1999, Bernoulli, 5, 3

\bibitem[{Andersson {et~al.}(2007)Andersson, Gustavsson, Mehlig, \&
  Wilkinson}]{andersson2007advective}
Andersson, B., Gustavsson, K., Mehlig, B., \& Wilkinson, M. 2007, EPL
  (Europhysics Letters), 80, 69001

\bibitem[{Armitage(2010)}]{armitage2010astrophysics}
Armitage, P.~J. 2010, Astrophysics of planet formation (Cambridge University
  Press)

\bibitem[{{Bec}(2003)}]{Bec03}
{Bec}, J. 2003, Physics of Fluids, 15, L81

\bibitem[{{Bec}(2005)}]{Bec05}
{Bec}, J. 2005, Journal of Fluid Mechanics, 528, 255

\bibitem[{{Bec} {et~al.}(2007{\natexlab{a}}){Bec}, {Biferale}, {Cencini},
  {Lanotte}, {Musacchio}, \& {Toschi}}]{Bec07b}
{Bec}, J., {Biferale}, L., {Cencini}, M., {et~al.} 2007{\natexlab{a}}, Phys.
  Rev. Letters, 98, 084502

\bibitem[{{Bec} {et~al.}(2007{\natexlab{b}}){Bec}, {Cencini}, \&
  {Hillerbrand}}]{Bec07}
{Bec}, J., {Cencini}, M., \& {Hillerbrand}, R. 2007{\natexlab{b}}, Phys. Rev.
  E, 75, 025301

\bibitem[{{Bhatnagar} {et~al.}(2018){Bhatnagar}, {Gustavsson}, \&
  {Mitra}}]{Bhatnagar18}
{Bhatnagar}, A., {Gustavsson}, K., \& {Mitra}, D. 2018, Phys. Rev. E, 97,
  023105

\bibitem[{Bird(1978)}]{bird1978monte}
Bird, G. 1978, Annu. Rev. Fluid Mech., 10, 11

\bibitem[{Bird(1981)}]{bird1981monte}
Bird, G. 1981, Progress in Astronautics and Aeronautics, 74, 239

\bibitem[{Birnstiel {et~al.}(2016)Birnstiel, Fang, \&
  Johansen}]{birnstiel2016dust}
Birnstiel, T., Fang, M., \& Johansen, A. 2016, Space Science Reviews, 205, 41

\bibitem[{{Bourdin}(2020)}]{2020GApFD}
{Bourdin}, P.-A. 2020, Geophys. Astrophys. Fluid Dynam., 114, 235

\bibitem[{Brandenburg(2001)}]{Brandenburg01}
Brandenburg, A. 2001, Astrophys. J., 550, 824

\bibitem[{Brandenburg \& Dobler(2002)}]{BD02}
Brandenburg, A. \& Dobler, W. 2002, Comput. Phys. Commun., 147, 471

\bibitem[{Brandenburg {et~al.}(2020)Brandenburg, Johansen, Bourdin, Dobler,
  Lyra, Rheinhardt, Bingert, Haugen, Mee, Gent,
  {et~al.}}]{brandenburg2020pencil}
Brandenburg, A., Johansen, A., Bourdin, P., {et~al.} 2020, arXiv preprint
  arXiv:2009.08231

\bibitem[{{Draine} \& {Salpeter}(1979)}]{Draine79}
{Draine}, B.~T. \& {Salpeter}, E.~E. 1979, ApJ, 231, 438

\bibitem[{Eaton \& Fessler(1994)}]{Eaton94}
Eaton, J. \& Fessler, J. 1994, International Journal of Multiphase Flow, 20,
  169

\bibitem[{Elmegreen \& Scalo(2004)}]{elmegreen2004interstellar}
Elmegreen, B.~G. \& Scalo, J. 2004, Annu. Rev. Astron. Astrophys., 42, 211

\bibitem[{Falkovich {et~al.}(2002)Falkovich, Fouxon, \&
  Stepanov}]{falkovich2002acceleration}
Falkovich, G., Fouxon, A., \& Stepanov, M. 2002, Nature, 419, 151

\bibitem[{Falkovich {et~al.}(2001)Falkovich, Gawedzki, \&
  Vergassola}]{falkovich2001particles}
Falkovich, G., Gawedzki, K., \& Vergassola, M. 2001, Reviews of modern Physics,
  73, 913

\bibitem[{{Federrath} {et~al.}(2009){Federrath}, {Klessen}, \&
  {Schmidt}}]{Federrath09}
{Federrath}, C., {Klessen}, R.~S., \& {Schmidt}, W. 2009, ApJ, 692, 364

\bibitem[{{Federrath} {et~al.}(2010){Federrath}, {Roman-Duval}, {Klessen},
  {Schmidt}, \& {Mac Low}}]{Federrath10}
{Federrath}, C., {Roman-Duval}, J., {Klessen}, R.~S., {Schmidt}, W., \& {Mac
  Low}, M.-M. 2010, A\&A, 512, A81

\bibitem[{Grabowski \& Wang(2013)}]{Grabowski_2013}
Grabowski, W.~W. \& Wang, L.-P. 2013, Annu. Rev. Fluid Mech., 45, 293

\bibitem[{Gustavsson \& Mehlig(2014)}]{Gustavsson13}
Gustavsson, K. \& Mehlig, B. 2014, J. Turbulence, 15, 34

\bibitem[{Gustavsson \& Mehlig(2016)}]{GM16}
Gustavsson, K. \& Mehlig, B. 2016, Advances in Physics, 65, 1

\bibitem[{Haugen {et~al.}(2004)Haugen, Brandenburg, \& Mee}]{haugen2004mach}
Haugen, N. E.~L., Brandenburg, A., \& Mee, A.~J. 2004, MNRAS, 353, 947

\bibitem[{{Hedvall} \& {Mattsson}(2019)}]{Hedvall19}
{Hedvall}, R. \& {Mattsson}, L. 2019, RNAAS, 3, 82

\bibitem[{{Hirashita}(2010)}]{Hirashita10}
{Hirashita}, H. 2010, MNRAS, 407, L49

\bibitem[{{Hirashita} {et~al.}(2014){Hirashita}, {Asano}, {Nozawa}, {Li}, \&
  {Liu}}]{Hirashita14}
{Hirashita}, H., {Asano}, R.~S., {Nozawa}, T., {Li}, Z.-Y., \& {Liu}, M.-C.
  2014, \planss, 100, 40

\bibitem[{{Hirashita} \& {Yan}(2009)}]{Hirashita09}
{Hirashita}, H. \& {Yan}, H. 2009, MNRAS, 394, 1061

\bibitem[{{Hopkins} \& {Lee}(2016)}]{Hopkins16}
{Hopkins}, P.~F. \& {Lee}, H. 2016, MNRAS, 456, 4174

\bibitem[{Johansen \& Lambrechts(2017)}]{johansen2017forming}
Johansen, A. \& Lambrechts, M. 2017, Annual Review of Earth and Planetary
  Sciences, 45, 359

\bibitem[{{Johansen} {et~al.}(2012){Johansen}, {Youdin}, \&
  {Lithwick}}]{Johansen_2012}
{Johansen}, A., {Youdin}, A.~N., \& {Lithwick}, Y. 2012, Astron. Astroph., 537,
  A125

\bibitem[{Jorgensen {et~al.}(1983)Jorgensen, Chandrasekhar, Madura, Impey, \&
  Klein}]{jorgensen1983comparison}
Jorgensen, W.~L., Chandrasekhar, J., Madura, J.~D., Impey, R.~W., \& Klein,
  M.~L. 1983, J. Chem. Phys., 79, 926

\bibitem[{Klett \& Davis(1973)}]{klett1973theoretical}
Klett, J. \& Davis, M. 1973, Journal of the Atmospheric Sciences, 30, 107

\bibitem[{{Kwok}(1975)}]{Kwok75}
{Kwok}, S. 1975, ApJ, 198, 583

\bibitem[{Li {et~al.}(2017)Li, Brandenburg, Haugen, \&
  Svensson}]{li2017eulerian}
Li, X.-Y., Brandenburg, A., Haugen, N. E.~L., \& Svensson, G. 2017, J. Adv.
  Modeling Earth Systems, 9, 1116

\bibitem[{Li {et~al.}(2020)Li, Brandenburg, Svensson, Haugen, Mehlig, \&
  Rogachevskii}]{Li19}
Li, X.-Y., Brandenburg, A., Svensson, G., {et~al.} 2020, Journal of the
  Atmospheric Sciences, 77, 337

\bibitem[{Li {et~al.}(2018)Li, Brandenburg, Svensson, Haugen, Mehlig, \&
  Rogachevskii}]{li2017effect}
Li, X.-Y., Brandenburg, A., Svensson, G., {et~al.} 2018, Journal of the
  Atmospheric Sciences, 75, 3469

\bibitem[{{Mattsson}(2011)}]{Mattsson11b}
{Mattsson}, L. 2011, MNRAS, 414, 781

\bibitem[{{Mattsson}(2016)}]{Mattsson16}
{Mattsson}, L. 2016, P\&SS, 133, 107

\bibitem[{{Mattsson} {et~al.}(2019){Mattsson}, {Bhatnagar}, {Gent}, \&
  {Villarroel}}]{Mattsson19a}
{Mattsson}, L., {Bhatnagar}, A., {Gent}, F.~A., \& {Villarroel}, B. 2019,
  \mnras, 483, 5623

\bibitem[{Mattsson {et~al.}(2019)Mattsson, Fynbo, \& Villarroel}]{Lars19_Small}
Mattsson, L., Fynbo, J. P.~U., \& Villarroel, B. 2019, MNRAS, 490, 5788

\bibitem[{{Mattsson} \& {Hedvall}(2021)}]{Mattsson21}
{Mattsson}, L. \& {Hedvall}, R. 2021, MNRAS, in prep.

\bibitem[{Maxey(1987)}]{Maxey87}
Maxey, M.~R. 1987, Journal of Fluid Mechanics, 174, 441–465

\bibitem[{{Pan} \& {Padoan}(2013)}]{Pan13}
{Pan}, L. \& {Padoan}, P. 2013, ApJ, 776, 12

\bibitem[{Pan \& Padoan(2014)}]{pan2014turbulence}
Pan, L. \& Padoan, P. 2014, ApJ, 797, 101

\bibitem[{Pan \& Padoan(2015)}]{pan2015turbulence}
Pan, L. \& Padoan, P. 2015, ApJ, 812, 10

\bibitem[{Pan {et~al.}(2014{\natexlab{a}})Pan, Padoan, \&
  Scalo}]{pan2014turbulenceII}
Pan, L., Padoan, P., \& Scalo, J. 2014{\natexlab{a}}, ApJ, 791, 48

\bibitem[{Pan {et~al.}(2014{\natexlab{b}})Pan, Padoan, \&
  Scalo}]{pan2014turbulenceIII}
Pan, L., Padoan, P., \& Scalo, J. 2014{\natexlab{b}}, ApJ, 792, 69

\bibitem[{{Pan} {et~al.}(2011){Pan}, {Padoan}, {Scalo}, {Kritsuk}, \&
  {Norman}}]{Pan11}
{Pan}, L., {Padoan}, P., {Scalo}, J., {Kritsuk}, A.~G., \& {Norman}, M.~L.
  2011, ApJ, 740, 6

\bibitem[{Pan {et~al.}(2011)Pan, Padoan, Scalo, Kritsuk, \&
  Norman}]{pan2011turbulent}
Pan, L., Padoan, P., Scalo, J., Kritsuk, A.~G., \& Norman, M.~L. 2011, The
  Astrophysical Journal, 740, 6

\bibitem[{{Passot} \& {V{\'a}zquez-Semadeni}(1998)}]{Passot98}
{Passot}, T. \& {V{\'a}zquez-Semadeni}, E. 1998, Phys. Rev. E, 58, 4501

\bibitem[{{Price} {et~al.}(2011){Price}, {Federrath}, \& {Brunt}}]{Price11}
{Price}, D.~J., {Federrath}, C., \& {Brunt}, C.~M. 2011, ApJL, 727, L21

\bibitem[{Procacia {et~al.}(1983)}]{procacia1983measuring}
Procacia, I. {et~al.} 1983, Physica. D, 9, 189

\bibitem[{Reade \& Collins(2000)}]{reade2000effect}
Reade, W.~C. \& Collins, L.~R. 2000, Physics of Fluids, 12, 2530

\bibitem[{{Rowlands} {et~al.}(2014){Rowlands}, {Gomez}, {Dunne},
  {Arag{\'o}n-Salamanca}, {Dye}, {Maddox}, {da Cunha}, \& {van der
  Werf}}]{Rowlands14}
{Rowlands}, K., {Gomez}, H.~L., {Dunne}, L., {et~al.} 2014, MNRAS, 441, 1040

\bibitem[{Saffman \& Turner(1956)}]{1955_Saffman}
Saffman, P.~G. \& Turner, J.~S. 1956, J. Fluid Mech., 1, 16

\bibitem[{{Schaaf}(1963)}]{Schaaf63}
{Schaaf}, S.~A. 1963, Handbuch der Physik, 3, 591

\bibitem[{{Smoluchowski}(1916)}]{Smoluchowski16}
{Smoluchowski}, M.~V. 1916, Zeitschrift fur Physik, 17, 557

\bibitem[{Squires \& Eaton(1991)}]{Squires91}
Squires, K.~D. \& Eaton, J.~K. 1991, Physics of Fluids A: Fluid Dynamics, 3,
  1169

\bibitem[{Sundaram \& Collins(1997)}]{Sundaram97}
Sundaram, S. \& Collins, L.~R. 1997, Journal of Fluid Mechanics, 335, 75–109

\bibitem[{{Valiante} {et~al.}(2011){Valiante}, {Schneider}, {Salvadori}, \&
  {Bianchi}}]{Valiante11}
{Valiante}, R., {Schneider}, R., {Salvadori}, S., \& {Bianchi}, S. 2011, MNRAS,
  416, 1916

\bibitem[{Vo{\ss}kuhle {et~al.}(2014)Vo{\ss}kuhle, Pumir, L{\'e}v{\^e}que, \&
  Wilkinson}]{vosskuhle2014prevalence}
Vo{\ss}kuhle, M., Pumir, A., L{\'e}v{\^e}que, E., \& Wilkinson, M. 2014,
  Journal of fluid mechanics, 749, 841

\bibitem[{{Wang} {et~al.}(2000){Wang}, {Wexler}, \& {Zhou}}]{Wang00}
{Wang}, L.-P., {Wexler}, A.~S., \& {Zhou}, Y. 2000, Journal of Fluid Mechanics,
  415, 117

\bibitem[{Wilkinson \& Mehlig(2005)}]{wilkinson2005caustics}
Wilkinson, M. \& Mehlig, B. 2005, EPL (Europhysics Letters), 71, 186

\bibitem[{Wilkinson {et~al.}(2006)Wilkinson, Mehlig, \& Bezuglyy}]{Wilkinson06}
Wilkinson, M., Mehlig, B., \& Bezuglyy, V. 2006, Phys. Rev. Lett., 97, 048501

\bibitem[{{Yavuz} {et~al.}(2018){Yavuz}, {Kunnen}, {van Heijst}, \&
  {Clercx}}]{Yavuz18}
{Yavuz}, M.~A., {Kunnen}, R.~P.~J., {van Heijst}, G.~J.~F., \& {Clercx},
  H.~J.~H. 2018, Physical Review Letters, 120, 244504

\bibitem[{{Zsom} \& {Dullemond}(2008)}]{Dullemond_2008}
{Zsom}, A. \& {Dullemond}, C.~P. 2008, Astron. Astrophys., 489, 931

\end{thebibliography}

\begin{appendix}
\section{Coagulation kernel: An alternative expression}
\label{app:kernel}
This appendix presents a discussion of an alternative description of the coagulation rate and the underlying mechanisms.
\Eq{kernel_gen} appears to be a natural expression for
the collision kernel between particle pairs. However, it is not
a physically complete and always representative model \citep{Wilkinson06}. Two mechanisms
contribute to the collision kernel $C_{ij}$ due to particle inertia: clustering and {\it \textup{caustics}}.
The former is characterised by $g(r)\propto r^{D_2-d}$ at a
fixed Stokes number, where $g$ is the radial distribution function, $d$ is the
spatial dimension, and $D_2$ is the correlation dimension \citep{reade2000effect,procacia1983measuring}.
The latter is the effect of singularities in the particle phase space at non-zero values of
the Stokes number \citep{falkovich2002acceleration,Wilkinson06,Gustavsson13}.
It appears when phase-space manifolds fold over. In the fold region, the velocity field
at a given point in space becomes multi-valued, allowing for
large velocity differences between nearby particles, which
results in a temporarily increased particle-interaction rate and more efficient coagulation \citep{Gustavsson13}.
As indicated above, the relative velocity between two colliding pairs obeys
a power law $|\langle w_{ij}\rangle|\propto (r_i+r_j)^{d-D_2}$.
Thus, the product of $g(r)$ and $|\langle w_{ij} \rangle|$
is independent of $(r_i+r_j)$, that is, caustics and
clustering cancel each other out in this formulation.
Therefore \citet{Wilkinson06} proposed that $C_{ij}$ is
a superposition of clustering and caustics, which was confirmed
by numerical simulations \citep{vosskuhle2014prevalence}.
In section~\ref{sec:theory} of our study we ignored the effects of caustics mainly because caustics in high $\mathcal{M}_{\rm rms}$ compressible turbulence are associated with shock interaction and density variance in the carrier fluid, in which case the resultant increase in particle number density is a far greater effect than the caustics.
\end{appendix}

\end{document}